\documentclass[12pt]{article}
\usepackage[T1]{fontenc}
\usepackage[latin9]{inputenc}
\usepackage{geometry}
\usepackage{graphicx}
\usepackage{esint}
\usepackage{subscript}

\makeatletter

\providecommand{\tabularnewline}{\\}

\usepackage{times,graphics}
\usepackage[authoryear]{natbib}
\newcommand{\rrh}{\widehat{\mbox{RR}}}
\usepackage[unicode=true]{hyperref}

\makeatother

\begin{document}

\title{Quantifying statistical uncertainty in the attribution of human influence
on severe weather}

\author{Christopher J. Paciorek\\Department of Statistics, University of California\\Berkeley, California, USA\\paciorek@stat.berkeley.edu \\ \\
Dáithí A. Stone\\Computational Research Division, Lawrence Berkeley National Laboratory\\Berkeley, California, USA\\dstone@lbl.gov \\ \\ 
Michael F. Wehner\\Computational Research Division, Lawrence Berkeley National Laboratory\\Berkeley, California, USA\\mfwehner@lbl.gov}

\maketitle
\newpage

\begin{abstract}
Event attribution in the context of climate change seeks to understand
the role of anthropogenic greenhouse gas emissions on extreme weather
events, either specific events or classes of events. A common approach
to event attribution uses climate model output under factual (real-world)
and counterfactual (world that might have been without anthropogenic
greenhouse gas emissions) scenarios to estimate the probabilities
of the event of interest under the two scenarios. Event attribution
is then quantified by the ratio of the two probabilities. While this
approach has been applied many times in the last 15 years, the statistical
techniques used to estimate the risk ratio based on climate model
ensembles have not drawn on the full set of methods available in the
statistical literature and have in some cases used and interpreted
the bootstrap method in non-standard ways. We present a precise frequentist
statistical framework for quantifying the effect of sampling uncertainty
on estimation of the risk ratio, propose the use of statistical methods
that are new to event attribution, and evaluate a variety of methods
using statistical simulations. We conclude that existing statistical
methods not yet in use for event attribution have several advantages
over the widely-used bootstrap, including better statistical performance
in repeated samples and robustness to small estimated probabilities.
Software for using the methods is available through the climextRemes
package available for R or Python. While we focus on frequentist statistical
methods, Bayesian methods are likely to be particularly useful when
considering sources of uncertainty beyond sampling uncertainty. 

\end{abstract}

\section{Introduction}

Over the past decade, there has been increasing interest in the climate
change research community in describing the role of anthropogenic
greenhouse gas emissions in specific extreme weather events, commonly
referred to as ``event attribution'' \\
\citep{Stot_etal_2013,NAS_2016,Herr_etal_2016}.
This increased scientific interest has been motivated by public interest,
as the global warming signal both becomes more noticeable and easier
to analyse, and an expectation that further understanding of the role
of anthropogenic emissions might inform adaptation activities. Concurrently,
observationally-based data products, numerical climate models, and
computational resources have developed to the point where they can
be usefully applied to the analysis of long-term trends in extreme
weather.

\citet{Alle_2003} first noted the potential public demand for event
attribution information and suggested that concepts from epidemiology
and environmental law would be appropriate in the climate change setting
as well. In this setting, one compares the probability of an extreme
weather event under a factual scenario of recent and current conditions
to the probability in a counterfactual scenario in which anthropogenic
emissions had never occurred but other factors (e.g., the eruption
of Mt. Pinatubo) had still influenced the climate system. \cite{Ston_Alle_2005}
formally introduced the concepts of fraction attributable risk (FAR)
and risk ratio (RR) and proposed how they might be estimated given
available tools. (Note that 'risk' in risk ratio inherits from its
usage in biostatistics/epidemiology and is unrelated to statistical
risk.) Unlike in epidemiology, in which repeated observations (e.g.,
multiple patients) are available with different exposures to potential
health risks, in the climate context we do not have available repeated
samples of the world, particularly under the counterfactual scenario.
Analyses therefore use simulations of numerical models of the climate
system as surrogates. \citet{Stot_etal_2004} presented the first
study using this approach, making use of a small number of simulations
of a climate model representing the entire climate system, but the
small number of simulations meant that they had to assume a relationship
between the average summer temperature and the frequency of extremely
hot summers. \citet{Pall_etal_2011} extended the method to take advantage
of an exceptionally large number of simulations that removed the need
for the assumption of a mean-extreme relationship, at the expense
of an incomplete model of the climate system. Together, these two
variants of the ``probabilistic event attribution'' method have
become the most popular approach for event attribution research in
recent years \citep[e.g.,][]{Pete_etal_2012,Pete_etal_2013,Herr_etal_2014,Herr_etal_2015,Herr_etal_2016}.

Despite this popularity, there has been little formal statistical
development of the technique since \citet{Ston_Alle_2005}. The
exception is the work of \citet{Hans_etal_2014}, who considered the
problem of estimating the FAR as the ratio of means of two independent
Poisson processes and proposed use of a confidence interval well-established
in the statistical literature \citep{Wils_1927}.  They also considered
the case when the estimated counterfactual probability of an event
is zero and developed a one-sided interval for FAR.

Underlying probabilistic event attribution are the existence of observational
data and relevant climate model simulations, along with the assumption
that both are adequate for performing an analysis. Guidelines for
such adequacy tests are still being considered \citep{Stot_etal_2004, Pall_etal_2011,Lott_Stot_2016,Ange_etal_2016}.
We leave the question of how to evaluate the input data sources for
further discussion elsewhere and focus instead on statistical methods
to estimate the RR (conditional on the data sources) and to quantify
the uncertainty in the estimate due to limited statistical sampling
because of finite model ensemble sizes. 

This paper presents a formal frequentist statistical framework for
probabilistic event attribution. Previous work has frequently involved
interpretation of frequentist-based analysis using a Bayesian perspective
\citep[e.g.,][]{Stot_etal_2004,Pall_etal_2011}, so part of our goal
is to clarify ways to implement standard frequentist statistical methods
in this context. We focus on estimating uncertainty from limited sampling,
for which a frequentist approach is well-suited. We present several
statistical methods for estimating uncertainty via confidence intervals,
assess performance in a statistical simulation study, provide an example
analysis, and make recommendations for the practice of probabilistic
event attribution. We highlight that some methods allow estimation
of a confidence interval even when the estimated counterfactual probability
is zero and the risk ratio is infinity.

\section{Statistical framework\label{sec:Statistical-framework}}

We consider the approach of estimating the risk (probability) ratio,
as the ratio of the probability of a specified event under a factual
scenario (F), $p_{F}$, to that probability under a counterfactual
scenario (CF), $p_{C}$:
\[
\mbox{RR}=\frac{p_{F}}{p_{C}}.
\]
In most climate attribution studies (and a case study in Section \ref{sec:example}),
the factual scenario is intended to represent the climate we have
experienced, i.e., in which historical emissions of greenhouse gases
and other drivers of climate change occurred as they did. In contrast,
the counterfactual is intended to represent a climate that might have
been in the absence of human interference, i.e., in which the anthropogenic
drivers are held at some ``pre-industrial'' level (e.g., year 1850
values) even as the natural drivers (volcanic eruptions and solar
luminosity) remain varying as we have experienced. Given estimates,
$\hat{p}_{F}$ for $p_{F}$ and $\hat{p}_{C}$ for $p_{C}$, one simply
estimates $\rrh=\frac{\hat{p}_{F}}{\hat{p}_{C}}$. We focus on RR
rather than FAR because of its use in epidemiology and related fields,
its interpretability, and because in our statistical development we
generally work with $\log(\mbox{RR)}$, which is expressed simply
as the difference in log probabilities and which improves statistical
performance. Usefully, on the log scale, increases and decreases from
anthropogenic influence are symmetric about zero, unlike for the FAR.
However, a confidence interval for FAR can be trivially calculated
from a confidence interval for RR.

In this work we lay out a classical (or frequentist) statistical framework
that treats $p_{F}$ and $p_{C}$ (and therefore RR) as fixed, non-random
quantities that are properties of the climate system. Uncertainty
arises in estimating these quantities, and our goal is to quantify
the uncertainty in $\rrh$ as an estimate of the unknown RR based
on the (sampling) probability density of $\rrh$. In contrast, a Bayesian
treatment would consider these quantities to be random variables with
probability densities and would allow one to make probabilistic statements
directly about RR. Note that in previous event attribution work, researchers
have frequently interpreted results, including bootstrap-based analyses,
in a Bayesian fashion despite largely relying on frequentist methods
\citep[e.g.,][]{Stot_etal_2004,Pall_etal_2011}.

\subsection{Basic probabilistic framework}

\label{sub:basic_framework}

We use $R$ to denote the continuous variable of interest (e.g., runoff
or rainfall) and $I(R>c)$ to denote the occurrence of the event of
interest, defined by whether the variable exceeds some cutoff, $c$.
$I(\cdot)$ is the indicator function that is 1 if the condition occurs
and 0 if not.

Let 
\begin{equation}
p=P(R>c)=E(I(R>c))=\int I(R>c)f(r)dr\label{eq:define-p}
\end{equation}
be the probability of the event, where $E(\cdot)$ denotes expected
value and $f(r)$ is the probability density function of $R$. While
in principle we can estimate this quantity in the real world using
observations, we only have a single observational series. If that
series were stationary (which may be a reasonable assumption only
for short time periods), we might use multiple years as replicates,
but we cannot estimate this quantity under the counterfactual scenario.
Therefore researchers rely on climate model simulations to estimate
$p\in\{p_{F},p_{C}\}$. (However the methods in this paper can be
used for a RR contrasting probabilities under two different time periods.)

In a single climate model simulation, $R$ for a pre-specified time
and location is not random since the model is deterministic and therefore
$R$ (and $I(R>c)$) is a fixed value. In this case, the use of expectation
and probability above is not meaningful. However, not only can models
be run under multiple scenarios, but they can be run multiple times
with each realization differing in the initial state and thus the
subsequent weather produced. Let $W$ (for ``weather'') be a (very
high-dimensional) random variable indicating the state of the earth
system and $f(w)$ the probability density function of $W$. The individual
realizations provide a simple random sample, $w_{1},\ldots,w_{n_{w}}$,
of size $n_{w}$ from $f(w)$. By using the initial condition ensemble,
passed through the deterministic GCM, to represent $f(w)$, we rely
on the strong assumption that use of the sample of initial conditions
to initialize the GCM can approximate the true $f(w)$ under a given
scenario. This ergodic assumption is reasonable for the atmospheric
model simulations used in the approach of \citet{Pall_etal_2011}
after a spin-up period of weeks to over a year but requires decades
for atmosphere-ocean GCMs.  $W$ is random and induces the randomness
in $R$. Thus $R(W)$ is our random variable of interest. Now we have
\begin{equation}
p=P(R(W)>c)=E(I(R(W)>c))=\int I(R(w)>c)f(w)dw.\label{eq:p-def}
\end{equation}

The discussion above assumes one value of $R$ per model realization,
which would often be the case for annual or seasonal extremes or for
short-term extremes for specific calendar dates. If one has multiple
values of $R$ for a given realization (e.g., analyzing daily rainfall),
one would generally benefit from having a much larger sample size
than is the focus here (particularly the simulation results of Section
\ref{sub:simulation}), but one would need to account for any correlation
between the multiple values of $R$ from a given realization (e.g.,
the daily values in a given season or year), particularly when estimating
uncertainty.

In Section \ref{sec:Estimating-event-probabilities} we present methods
to estimate $p\in\{p_{F},p_{C}\}$, while in Section \ref{sec:Estimating-uncertainty-in}
we discuss how to quantify the uncertainty in estimating RR. Before
doing so, we consider the additional complexities involved in estimating
RR that are not characterized in the basic framework above.

\subsection{Sources of uncertainty\label{sub:Sources-of-uncertainty}}

Sources of uncertainty in estimating the RR using GCMs (see also \\\citet{NAS_2016})
include: 
\begin{itemize}
\item \textbf{variability in the earth system:} Uncertainty arises from
limited sampling of the variability of the system, which can range
from time scales of days to years, decades, and centuries. This is
often operationalized via an ensemble of GCM simulations initialized
with different initial conditions and quantified based on the notion
of sampling uncertainty as discussed in the basic framework above.
\item \textbf{boundary condition uncertainty}: This includes aspects of
the system that can vary in time but that are prescribed in the model
and thus not simulated by it, such as the concentrations of radiatively-active
trace constituents in the atmosphere and changes in land use. Modification
of one or more of these boundary conditions constitutes the distinction
between the factual and counterfactual scenarios; thus uncertainty
in the distinction is implicit in this source of uncertainty. When
such forcing terms are based on observational data, there is observational
uncertainty.
\item \textbf{model parametric uncertainty: }This represents uncertainty
in the appropriate values for parameters in the GCM that are used
in approximations for various processes not directly simulated by
the GCM. This could include fundamental physical or chemical constants,
but generally uncertainty in those constants is negligible relative
to uncertainty in the appropriate value of bespoke parameters.
\item \textbf{model structural uncertainty:} This is another component of
uncertainty inherent to the climate model -- the uncertainty in how
to represent a complex physical system as a mathematical model, but
not the uncertainty in tuning of that approximation. Note that determination
of whether a model is fit for purpose for event attribution is a binary
determination of this broader component of uncertainty.
\end{itemize}
We draw a fundamental distinction between uncertainty arising from
variability in the system and the remaining sources of uncertainty.
Uncertainty from variability is a statistical sampling problem, with
uncertainty decreasing with increasing sample size. Critically, this
uncertainty is quantifiable using well-established frequentist or
Bayesian statistical methods. If we had an infinitely large ensemble,
our sampling uncertainty would be zero. However, there would still
be bias from the fact that the climate produced by the model is not
the same as the climate produced by the real system, even in equilibrium,
due to the additional factors listed above. In this work, we focus
on uncertainty from variability, in part because frequentist methods
would appear to be of limited use for the other sources. We make some
comments on these other components of uncertainty in the discussion. 

Finally, many event attribution analyses use atmospheric-land GCMs,
as in \citet{Pall_etal_2011}, rather than coupled models, for reasons
of computational efficiency, a partial conversion of (oceanic) model
structural uncertainty to better-understood boundary condition uncertainty,
and in some cases a desire to more strongly condition the analysis
on known features of the real world. With such analyses, model simulations
do not sample from the longer-time-scale internal variability of the
system. In Supplemental Material B  we describe the additional uncertainty
related to this longer-time-scale variability and present statistical
methods for averaging over results from multiple years.

\section{Estimating event probabilities \label{sec:Estimating-event-probabilities}}

We next provide an overview and comparison of methods for estimating
$p\in\{p_{F},p_{C}\}$. We discuss approaches that simply count exceedances
and rely on binomial sampling statistics, parametric fitting of the
variable of interest, and extreme value analysis (EVA) of the variable
of interest.

While EVA is often used for estimating probabilities of extreme events
with observational data, serious difficulties can arise in the context
of event attribution analyses that use model ensembles, and simple
nonparametric estimators are often a good choice. First note that
EVA is generally used on long time series and applied to short-term
(e.g., daily) extremes. In this context, one often uses a block maximum
(or minimum) approach, blocking by year, or a peaks-over-threshold
approach that only uses observations over a high threshold, such as
the 99th percentile of the observations. The statistical theory that
supports EVA relies on taking maxima (or minima) over a large number
of observations per block or setting a high threshold. However, for
analysis of annual or seasonal extremes or short-term extremes for
specific calendar dates, model ensembles generally only provide moderate
sample sizes such as 50, 100, or 400, violating the assumptions of
EVA. Furthermore, in event attribution, the event of interest may
be extreme only in one scenario, so EVA may not be applicable for
one scenario. And for short model simulations with initial conditions
that favor extreme weather, the entire sample may be extreme in an
absolute sense, in which case the event of interest may, in a relative
(i.e., conditional) sense, not be extreme and EVA would not be appropriate
\citep[e.g.,][]{Pall_etal_2017}. 

Thus, while there are situations in which EVA has advantages for model-based
event attribution (as seen in the case study in Section \ref{sec:example}
and discussed further in our recommendations in Section \ref{sec:discussion}),
we focus more on the binomial approach because of its greater generality.

\subsection{Nonparametric (binomial sampling of events)}

\label{par:nonparam}

One straightforward approach to estimating $p$ is a nonparametric
Monte Carlo (MC) estimate based on the available sample. An MC estimate
of the expectation in (\ref{eq:p-def}) involves drawing $n_{w}$
samples of $W$ from $f(w)$ (based on the model simulations) and
using the estimator:
\begin{equation}
\hat{p}=\frac{\mbox{\#events}}{n_{w}}=\frac{1}{n_{w}}\sum_{i=1}^{n_{w}}I(R(w_{i})>c),\label{eq:MC_estimator}
\end{equation}
which is justified by the law of large numbers because $\hat{p}$
is a statistically-consistent estimator for $p$ as $n_{w}\to\infty$.
Consistency here means that as $n_{w}$ gets large, $\hat{p}$ is
guaranteed to get very close to $p$. 

This estimator makes no assumptions about the distribution of $R$
(hence the term 'nonparametric', although it is equivalent to Bernoulli
sampling and a resulting binomial distribution for the number of events)
and is thus robust, but there are drawbacks to the approach. First,
the estimator may have more uncertainty than parametric estimators
that assume a particular distribution. Furthermore, if $R>c$ does
not occur in the sample, our estimator is $\hat{p}=0$ even when we
have substantive expertise suggesting that $p$ is non-zero. Zeros
for $\hat{p}_{F}$, $\hat{p}_{C}$, or both result in RR estimates
of zero, infinity, or an undefined value.

\subsection{Parametric}

If one is willing to assume a particular parametric form for the distribution
of $R$, such as a normal or log-normal distribution, statistical
theory tells us that one may be able to obtain an estimator for $p$
that has lower variance (less uncertainty) than the nonparametric
estimator above. For example if we assume normality, we can estimate
the mean and variance of $R$ as $\bar{r}$ and $s_{R}^{2}=\frac{\sum(r_{i}-\bar{r})^{2}}{n_{w}-1}$,
where $r_{i}=R(w_{i})$ and $\bar{r}$ is the simple average of $r_{1},\ldots,r_{n_{w}}$.
Then $\hat{p}=\int_{c}^{\infty}f(r;\bar{r},s_{R}^{2})dr$ where $f(r;\bar{r},s_{R}^{2})$
is the normal density with mean $\bar{r}$ and variance $s_{R}^{2}$.

The methodology allows us to estimate probabilities for events far
in the tail of the distribution, but it relies crucially on the assumption
that the chosen distribution well approximates the true distribution
even far in the tail. Put another way, all the data values are used
to estimate the parameters of the assumed distribution and thereby
infer the behavior of the tail of the distribution. 

Finally, note that unless the chosen distribution has a finite bound
(either a maximum for extremes in the upper tail or a minimum for
the lower tail), one will not obtain $\hat{p}=0$.

\subsection{Extreme value analysis (EVA)\label{sub:Extreme-value-analysis}}

A compromise between the fully parametric and nonparametric approaches
is to use extreme value techniques developed in the statistical literature;
\citet{Cole_2001} provides an excellent overview. This approach applies
when the event of interest is sufficiently far in the tail of the
distribution of $R$. The approach involves fitting a three-parameter
statistical distribution only to the extreme observations. While the
approach takes a parametric form, statistical theory provides strong
theoretical support for the particular distributional form. We provide
an overview of these methods, including a discussion of the statistical
and implementational challenges of using the methods for annual or
seasonal extremes in Supplemental Material C .

\section{Estimating uncertainty in the risk ratio\label{sec:Estimating-uncertainty-in}}

\label{sec:uncertainty}

In this section we consider several methods for estimating uncertainty
using frequentist confidence intervals. Our focus is on confidence
intervals for RR, and we introduce each method in the simplest context
of the nonparametric estimator (Section \ref{par:nonparam}) before
describing how the method could be used in the context of EVA. After
describing the methods we carry out a simulation study to assess which
performs best.

\subsection{The frequentist interpretation of uncertainty about $p$ and RR.}

As discussed briefly in Section \ref{sec:Statistical-framework},
we \emph{cannot} generate a distribution of $p$ or of $\mbox{RR}=p_{F}/p_{C}$
using the methods discussed here, as this is not part of the frequentist
statistical framework used here. $\rrh$ has a distribution, called
the sampling distribution, that is induced by the distributions of
$\hat{p}_{F}$ and $\hat{p}_{C}$, but it is not particularly useful
to plot it, as it represents the variability of $\rrh$ around RR.
The danger in presenting such a plot is that it will often be viewed
incorrectly as representing the distribution of RR, interpreted in
a Bayesian fashion even though it has not been derived in a Bayesian
fashion. For example, suppose the sampling distribution has a long
right tail. This indicates that the estimator, $\rrh$, might be much
larger than the true value, RR, and suggests that the true value may
be much smaller than the our estimate. But if this sampling distribution
were incorrectly interpreted as a Bayesian posterior, one would conclude
that the true value may be much larger than the point estimate. (However,
in the case of a simple parametric model and large sample size, the
Bayesian central limit theorem states that the Bayesian posterior
distribution will be approximately the same as the frequentist sampling
distribution \citep{Gelm_etal_2013}).

Instead, the standard statistical approach to quantifying uncertainty
is to estimate a confidence interval. For the sake of illustration
we will discuss a 90\% confidence interval here, but the ideas are
the same for other levels of confidence. The basic principle of a
90\% confidence interval is that it is a random interval that should
include the true value of the RR in 90\% of the datasets that we might
observe/collect. It is a random interval because it varies depending
on the data collected, while the true RR is considered to be fixed.
Therefore given a procedure for producing a confidence interval we
can assess the procedure based on its coverage probability (the probability
it contains the true value), assessed using synthetic datasets developed
to mimic the real data-generating mechanism. Our goal is to use a
procedure for calculating a confidence interval that produces intervals
as short as possible while still having the specified coverage probability
of 90\%.

\subsection{Methods for calculating confidence intervals}

In this section, we explore several ways to estimate a confidence
interval for the RR. We note that the statistical literature on estimating
the RR (and the related odds ratio) in the biomedical literature is
large and in particular considers a number of methods applied to binomial
counts (see \citet{Fage_etal_2015} for an overview), and we consider
some of those methods in addition to the bootstrap and a basic normal-theory
method. Many of these methods are implemented in the climextRemes
package available for R and Python, which builds upon the extRemes
R package \citep{Gill_Katz_2011}.

In general for probabilities (especially those near 0 or 1) and for
ratios the sampling distribution of a given estimator is not a normal
distribution, often being skewed. In this setting, working on the
log scale often gives better statistical performance, with confidence
intervals that come closer to having the desired coverage \citep[e.g.,][]{Katz_etal_1978}.
One would compute the confidence interval for the quantity on the
(natural) log scale (e.g., the log risk ratio) and then exponentiate
the endpoints of the interval to get a confidence interval on the
original scale.

\subsubsection{Normal-theory confidence intervals\label{sub:Normal-theory-confidence-interva}}

Basic statistical theory provides a standard confidence interval based
on the standard error of the estimator when the sampling distribution
of the estimator is approximately normally distributed. The approach
is appropriate for larger sample sizes but can perform poorly for
smaller sample sizes and when either $p_{F}$ or $p_{C}$ is close
to zero. Furthermore, it involves some mathematical derivation, with
use of the delta method in the context of the RR. For completeness
we provide details in Supplemental Material D.1 . 

We turn next to the bootstrap, which aims to overcome some of these
difficulties by relying on computation.

\subsubsection{Bootstrap\label{sub:Bootstrap}}

The bootstrap is a widely-used, asymptotically-justified statistical
tool for estimating the uncertainty in statistical estimates, particularly
in cases where one cannot derive the standard error of or confidence
interval for an estimator in closed form \citep{Efro_Tibs_1994,Davi_Hink_1997}.
To introduce the bootstrap we present it to develop a confidence interval
for $p$ for simplicity. However the same techniques work on the log
scale for $\log p$ and for $\log$RR.

A common version of the bootstrap involves resampling values of $R$
with replacement from the sample, $r_{1},\ldots,r_{n_{w}}$ to generate
$n_{b}$ different bootstrap datasets, each of size $n_{w}$. With
each bootstrap dataset we estimate $p$, giving us $\hat{p}^{(1)},\ldots,\hat{p}^{(n_{b})}$.
Note that the method does \emph{not} provide us with a distribution
to represent uncertainty in $p$ and should not be interpreted as
a Bayesian posterior distribution for $p$ (though there are connections
between Bayesian methods and the bootstrap). Rather the sample $\hat{p}^{(1)},\ldots,\hat{p}^{(n_{b})}$
provides us with an estimate of the sampling distribution of $\hat{p}$.
The idea of the bootstrap is that the empirical distribution of the
$\hat{p}^{(i)}$ values around $\hat{p}$ generally behaves similarly
to the distribution of $\hat{p}$ around $p$, which is the true sampling
distribution, $f(\hat{p};p)$. If we knew the true sampling distribution,
it would be simple to calculate a confidence interval. E.g., with
normal data, we know that $\bar{x}\sim\mbox{N}(\mu,\frac{\sigma^{2}}{n})$.
Given knowledge of this sampling distribution, we can analytically
derive a 90\% confidence interval for $\mu$ as $\bar{x}$ plus/minus
1.64 times the standard error, $\sigma/\sqrt{n}$, without needing
to use the bootstrap. In the absence of a known sampling distribution,
there are a variety of ways to use the $\hat{p}^{(i)}$ values to
estimate a bootstrap confidence interval for $p$:
\begin{itemize}
\item \textbf{Bootstrap interval using bootstrap standard error estimate}:
use the empirical standard deviation of the $\hat{p}^{(i)}$ values
around their mean, $\bar{\hat{p}}$. If we are happy to assume normality,
we can use the empirical standard deviation of the $\hat{p}^{(i)}$
values as the standard error estimate, which we denote $\widehat{se}$,
to form a 90\% confidence interval in the usual way,
\begin{equation}
\left(\hat{p}-1.64\cdot\widehat{se},\hat{p}+1.64\cdot\widehat{se}\right).\label{eq:normal_approx_boot_CI}
\end{equation}

\item \textbf{Bootstrap percentile interval}: use percentiles of the empirical
distribution of the $\hat{p}^{(i)}$ values to avoid the assumption
of normality needed when using the bootstrap-based standard error
estimate. This gives the following 90\% interval, involving the 5th
($\hat{p}^{5\%}$) and 95th ($\hat{p}^{95\%}$) percentiles of the
$\hat{p}^{(i)}$ values, 
\begin{equation}
\left(\hat{p}^{5\%},\hat{p}^{95\%}\right).\label{eq:percentile_CI}
\end{equation}

\item \textbf{Basic bootstrap interval}: the interval also involves $\hat{p}^{5\%}$
and $\hat{p}^{95\%}$ and is 
\begin{equation}
\left(\hat{p}-(\hat{p}^{95\%}-\hat{p}),\,\hat{p}-(\hat{p}^{5\%}-\hat{p})\right).\label{eq:basic_boot_CI}
\end{equation}

It is useful to consider the intuition behind why the upper quantile
is involved in calculating the lower endpoint of the interval and
vice versa, in contrast to the percentile method. The basic intuition
is that if $\hat{p}^{95\%}$ is much larger than $\hat{p}$, this
indicates the plausibility of estimated values that are much larger
than the true value. Given this, our lower limit for $p$ should be
much lower than $\hat{p}$, because our actual estimate, $\hat{p}$,
calculated using our single dataset may be much larger than the true
value, $p$. Note that if the sampling distribution is symmetric,
with $\hat{p}-\hat{p}^{5\%}=\hat{p}^{95\%}-\hat{p}$ then the percentile
and basic intervals will be the same.

\item \textbf{Studentized bootstrap interval}: this approach improves upon
the basic interval by standardizing by an estimate of the standard
error in each bootstrap sample. Let $z^{(i)}=(\hat{p}^{(i)}-\hat{p})/\widehat{se}^{(i)}$
where $\widehat{se}^{(i)}$ is a (possibly rough) estimate of the
standard error of $\hat{p}^{(i)}$ that is calculated from the $i$th
bootstrap sample. Let $z^{5\%}$ and $z^{95\%}$ be the 5th and 95th
percentiles of the $z^{(i)}$ values. The studentized confidence interval
is then
\[
\left(\hat{p}-\widehat{se}\cdot z^{95\%},\,\hat{p}-\widehat{se}\cdot z^{5\%}\right)
\]
where $\widehat{se}$ is the (possibly rough) standard error estimate
based on the actual dataset. 
\item \textbf{Adjusted percentile (BC\textsubscript{a}) bootstrap interval}:
this approach seeks to improve upon the percentile interval by estimating
a transformation that brings the sampling distribution closer to normality
\citep{Davi_Hink_1997}. 
\end{itemize}

\paragraph{Confidence intervals for RR}

In general, the ensemble members under the two scenarios are unrelated.
If they were related, we would want to resample from the data in a
way that reflected the relationship between the simulations in the
two scenarios. Given the lack of relationship, a bootstrap procedure
is to obtain $n_{b}$ resampled datasets from each scenario and to
randomly pair the datasets to get $n_{b}$ pairs, from which one can
calculate $\log\rrh^{(1)},\ldots,\log\rrh^{(n_{b})}$. This pertains
to estimates obtained from any of the nonparametric, parametric, or
EVA methods.

\paragraph{Drawbacks to the bootstrap }

An important difficulty with the bootstrap occurs when $\rrh=\infty$
(the following discussion holds equivalently for $\rrh=0)$. In this
case in which $\hat{p}_{C}=0$ the bootstrap fails because there is
no variability in the data; all bootstrap datasets will have $\hat{p}_{C}^{(i)}=0$.
Another case is when some of the bootstrap samples have $\hat{p}_{C}^{(i)}=0$
or $\hat{p}_{F}^{(i)}=0$ and therefore $\log\rrh^{(i)}\in\{-\infty,\infty\}$.
In this case, the bootstrap standard error estimate (and therefore
the bootstrap normal interval) cannot be calculated, while the ad
hoc approach of removing such bootstrap samples before calculating
confidence intervals using the various bootstrap methods has no clear
justification and could affect performance of the resulting confidence
interval. Finally when using EVA, it is not uncommon to be unable
to estimate $\hat{p}_{F}$ or $\hat{p}_{C}$ because the optimization
for the EVA parameters does not converge, and this often occurs with
a subset of the bootstrapped datasets as well. It is unclear how to
handle this, although if the number of times this occurs is small,
ad hoc calculation of confidence intervals based on ignoring these
values may not cause serious bias in the uncertainty quantification. 

More generally than just in this context of estimating the RR, the
percentile bootstrap method is known to perform poorly in practice
\citep{Hest_2015}. However, while the logic behind the swapping of
the percentiles in the basic bootstrap method relative to the percentile
method suggests we might favor the basic bootstrap, both theoretical
results \citep{Davi_Hink_1997,Hest_2015} and our simulation results
(Section \ref{sub:simulation}) show that the basic bootstrap also
does not perform well. Furthermore, our simulation results suggest
that all of the bootstrap-based methods have serious drawbacks. 

This is not surprising, particularly for values of $\hat{p}_{F}$
or $\hat{p}_{C}$ near zero. The bootstrap is known to perform poorly
when the resampling gives a discrete distribution with few values.
An extreme manifestation of this occurs when $\hat{p}_{N}$ or $\hat{p}_{A}$
are zero. However, even when this does not occur, the sampling distribution
often has probability mass at $\infty$, $0$ and $0/0$, and the
bootstrap estimate of the sampling distribution of $\log\rrh-\log\mbox{RR}$
can provide a poor approximation to the true sampling distribution.

\subsubsection{Inverting a hypothesis test using the likelihood ratio statistic\label{sub:Inverting-a-hypothesis}}

This approach is appealing because it can be used when $\hat{p}_{F}=0$
(or $\hat{p}_{C}=0$), providing a one-sided confidence interval that
gives a lower (upper) bound on plausible values of RR, as done in
\citet{Jeon_etal_2016}. The usefulness of providing a bound on the
RR should be apparent, because the bound allows us to assess the magnitude
of the anthropogenic influence in light of uncertainty. The basic
idea is to find a confidence interval by inverting a hypothesis test
for the estimate of interest. A standard hypothesis test that is commonly
applicable is a likelihood ratio test \citep{Case_Berg_2002}, which
compares the likelihood of the data based on the maximum likelihood
estimate to the likelihood of the data when constraining the parameters
to represent a simpler setting in which a null hypothesis is assumed
true. Confidence intervals based on likelihood ratio tests are widely
used and well described in the statistical literature but have not
been used in event attribution, so we provide details in Supplemental
Material D.2.

\subsubsection{Binomial-based intervals}

\label{sub:binom-test}

For the case where the two probabilities in the risk ratio are estimated
using the nonparametric approach, but not the EVA approach, the statistical
literature provides a wide variety of methods to calculate a confidence
interval for the risk ratio from independent binomial proportions,
motivated by the vast number of analyses of risk ratios in the biomedical
literature \citep{Fage_etal_2015}. Once again, a standard framework
involves inverting a hypothesis test. Of the wide array of possibilities,
some methods our literature search suggested to be promising are:
\begin{itemize}
\item Wilson's method: \citet{Hans_etal_2014} propose to find an interval
for the risk ratio by conditioning on the sum of events in the two
scenarios to produce a binomially-distributed quantity and using an
approximate confidence interval for a binomial probability proposed
by \citep{Wils_1927}.
\item Koopman's asymptotic score method: \citep{Koop_1984} proposes to
invert Pearson's chi square test. \citep{Fage_etal_2015} found this
method to perform reasonably well in simulations.
\item Wang and Shan's method: \citet{Wang_Shan_2015} seek to improve upon
existing methods by an inductive process to determine the ordering
of how extreme different data values are with respect to providing
evidence against the null hypothesis and then computing p-values by
computing probabilities of data as or more extreme than observed,
using the worst case value of the unknown nuisance parameter. One
can then invert the test to obtain a confidence interval. More details
are in Supplemental Material D.3. 
\end{itemize}

\section{Simulation study to evaluate the methods\label{sub:simulation}}

Given the variety of methods available, the potential for small sample
sizes, the fact that $p_{F}$ or $p_{C}$ are often near zero, and
the difficulty that the normality-based method and the bootstrap methods
have when $\rrh=\infty$, we explored the performance of the methods
using a simulation study. For simplicity and given the limitations
of EVA discussed earlier, the simulation study focuses on the context
of estimating the RR based on nonparametric approach rather than EVA.
The key factors that affect the statistical performance of the methods
are the ensemble size, the true value of the RR, and the true probability
of the event.

\subsection{Design}

For a given scenario, defined by the ensemble size ($n$), the RR,
and the event probability ($p_{F}$), we generated 5000 synthetic
datasets of $y_{F}$ and $y_{C}$ values, each from a binomial distribution
with $n$ ensemble members and probability of event $p_{F}$ and $p_{C}=p_{F}/\mbox{RR}$,
respectively. We considered $n\in\{25,50,100,400\}$, where $50$
is the minimum suggested in the C20C+ Detection and Attribution project
protocol (\url{http://portal.nersc.gov/c20c})  and 400 the number
available for our case study. Our results focus on ensembles of size
100 as intermediate between the high uncertainty with smaller sizes
and the fact that sizes as large as 400 will often not be available.
We considered $\mbox{RR}\in\{1,2,4,8,16\}$ representing the range
from a world where there is no anthropogenic influence to one with
very strong influence. We considered $p_{F}\in\{0.01,0.025,0.05,0.10,0.20\}$
to represent a range in how extreme the event is. Given all possible
combinations of values of $n$, RR, and $p_{F}$, we have 100 scenarios.

Note that while we only consider risk ratios greater than or equal
to one, the results are equivalent for $p_{C}\in\{0.01,0.025,0.05,0.10,0.20\}$
with $\mbox{RR}\in\{1,1/2,1/4,1/8,1/16\}$ because the same calculations
can be done for $\mbox{RR}^{-1}\equiv\frac{p_{C}}{p_{F}}$, followed
by taking one over the resulting interval endpoints to get a confidence
interval for RR. 

We then applied the following methods to each synthetic dataset, thereby
producing a set of 5000 confidence intervals for each method for each
scenario. The methods were:
\begin{itemize}
\item Wilson's method (Section \ref{sub:binom-test}),
\item Koopman's asymptotic score test inversion (Section \ref{sub:binom-test}),
\item Wang-Shan exact test inversion (Section \ref{sub:binom-test}),
\item normal-theory with delta method (Section \ref{sub:Normal-theory-confidence-interva}),
\item likelihood-ratio (LR) test inversion (Section \ref{sub:Inverting-a-hypothesis}),
\item bootstrap normal (Section \ref{sub:Bootstrap}),
\item percentile bootstrap (Section \ref{sub:Bootstrap}),
\item basic bootstrap (Section \ref{sub:Bootstrap}),
\item bootstrap-t (Section \ref{sub:Bootstrap}), and
\item BC\textsubscript{a} bootstrap (Section \ref{sub:Bootstrap}).
\end{itemize}
We report results for 90\% confidence intervals, considering the lower
and upper endpoints separately because in real event attribution analyses,
the focus will often be on the lower bound for the RR so as to assess
the plausibility of no anthropogenic influence. Thus we focus on 95\%
one-sided confidence intervals. However we note that some of the methods
were not developed as two separate one-sided intervals. Therefore,
a two-sided interval may cover the true RR with the correct probability
but when interpreted as two separate one-sided intervals not have
the correct one-sided coverage probability.

We assessed performance using the following metrics:
\begin{itemize}
\item coverage probability of the intervals (proportion of times the 5000
intervals included the true RR),
\item length of the interval (focusing on the magnitude of the lower bound),
and
\item proportion of intervals that could not be calculated because $\hat{p}_{F}=0$
or $\hat{p}_{C}=0$.
\end{itemize}
All code is available at \\\url{https://bitbucket.org/lbl-cascade/event_attribution_uq_paper}.

\subsection{Results}

\begin{figure}
\includegraphics{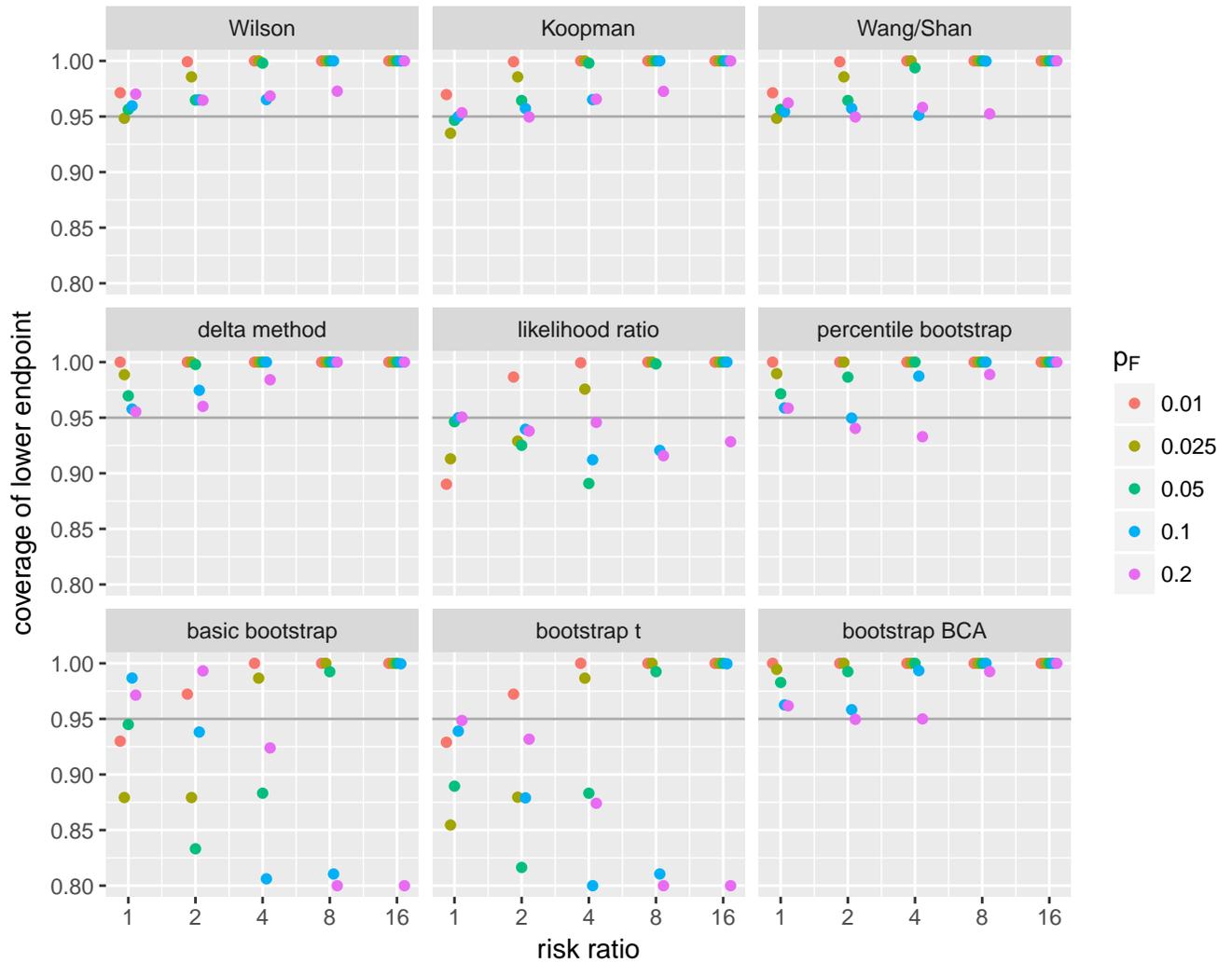}

\caption{Coverage probability of 95\% lower confidence bound for $n=100$ for
various methods and values of RR and $p_{F}$. Values of 0.95 are
optimal, while values less than 0.95 indicate undercoverage and values
greater than 0.95 indicate conservativeness (overcoverage). Values
lower than 0.8 are set to 0.8 for display purposes. Simulations in
which the lower bound could not be computed were excluded.}
\label{fig:coverage_low}
\end{figure}

Figure \ref{fig:coverage_low} shows the coverage probability for
a one-sided lower confidence interval, $(\mbox{RR}_{L},\infty)$.
We see that for the LR, basic bootstrap, and bootstrap-t intervals,
the intervals fail to include the true value at least 95\% of the
time. In contrast, the other methods are overly conservative (the
intervals include the true value more than 95\% of the time) for most
or all values of RR. While values higher than 95\% may sound appealing,
they provide conservative results with intervals that are overly long
and thereby increased uncertainty in estimating the RR.

Note that in this and other figures we omit results for the bootstrap
normal interval. In almost all of the scenarios the standard deviation
of the RR estimates in the bootstrap samples could not be calculated
because one or more estimates were zero, infinity or 0/0, and omitting
those estimates has no statistical justification.

\begin{figure}
\includegraphics{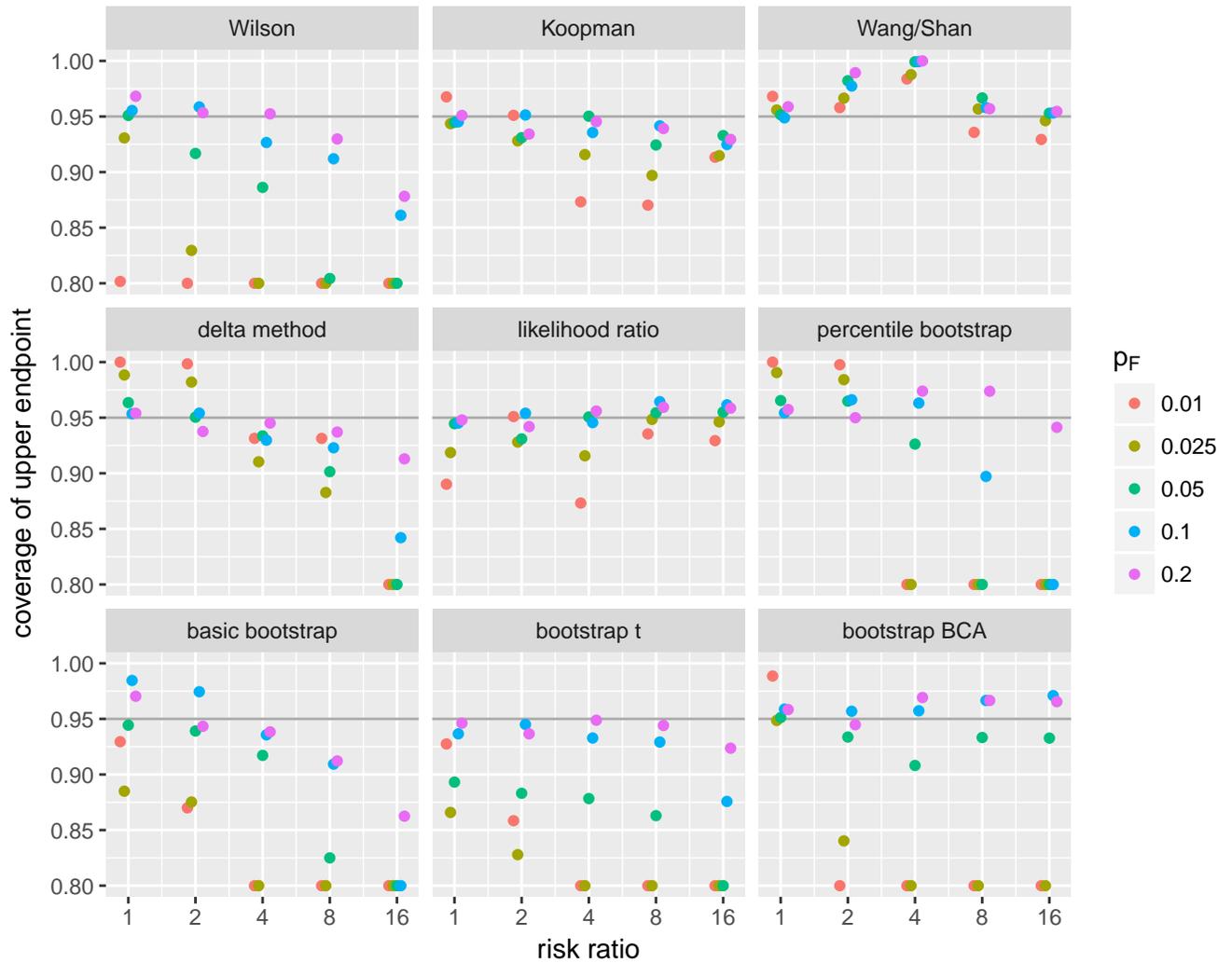}

\caption{Coverage probability of 95\% upper confidence bound for $n=100$ for
various methods and values of RR and $p_{F}$. Values of 0.95 are
optimal, while values less than 0.95 indicate undercoverage and values
greater than 0.95 indicate conservativeness (overcoverage). Values
lower than 0.8 are set to 0.8 for display purposes. Simulations in
which the upper bound could not be computed were excluded.}
\label{fig:coverage_high}
\end{figure}

Figure \ref{fig:coverage_high} shows the coverage probability for
a one-sided upper confidence interval, $(0,\mbox{RR}_{U})$. Here
the Wang-Shan method is conservative, the LR and Koopman methods show
some undercoverage, and the bootstrap, Wilson, and delta methods have
widely variable results, with substantial undercoverage in some cases. 

\begin{figure}
\includegraphics{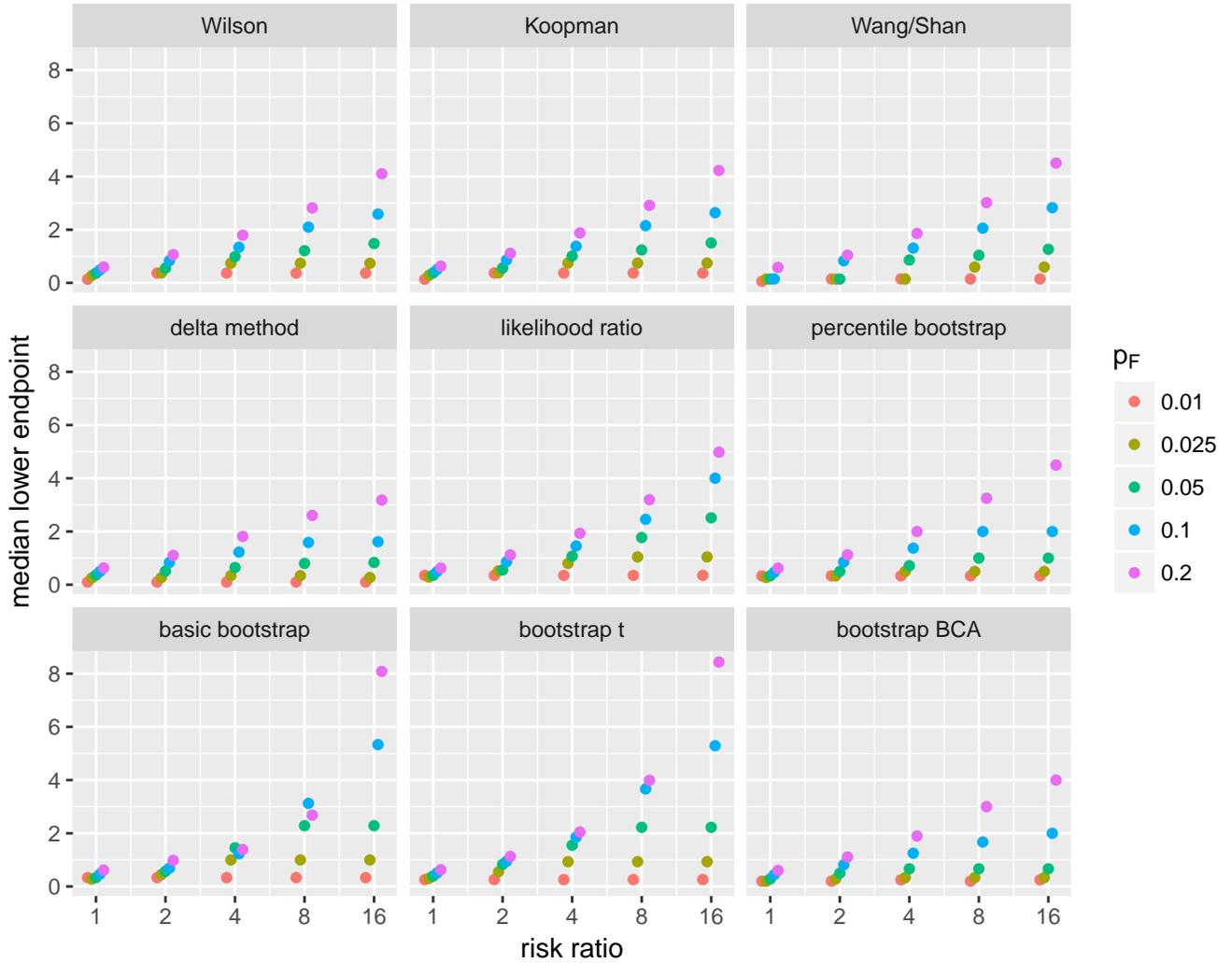}

\caption{Median value of lower confidence bound for $n=100$ for various methods
and values of RR and $p_{F}$. Higher values are better as they correspond
to shorter confidence intervals. Simulations in which the lower bound
could not be computed were excluded.}
\label{fig:value_low}
\end{figure}

Figure \ref{fig:value_low} shows the median value of $\mbox{RR}_{L}$,
with higher values corresponding to shorter intervals and more statistical
certainty about the value of RR. As expected, in general intervals
are wider for smaller values of $p_{F}$ (and therefore also of $p_{C})$,
corresponding to fewer observed events. While the LR, basic bootstrap,
and bootstrap-t methods have higher lower bounds, these are the methods
showing undercoverage (particularly the bootstrap methods), so the
shorter intervals are only obtained by violating the condition that
coverage probability be correct. In contrast, the Koopman, Wilson,
and Wang-Shan methods perform well, with the Koopman and Wilson approaches
showing shorter intervals for smaller values of $p_{F}$. 

\begin{figure}
\includegraphics{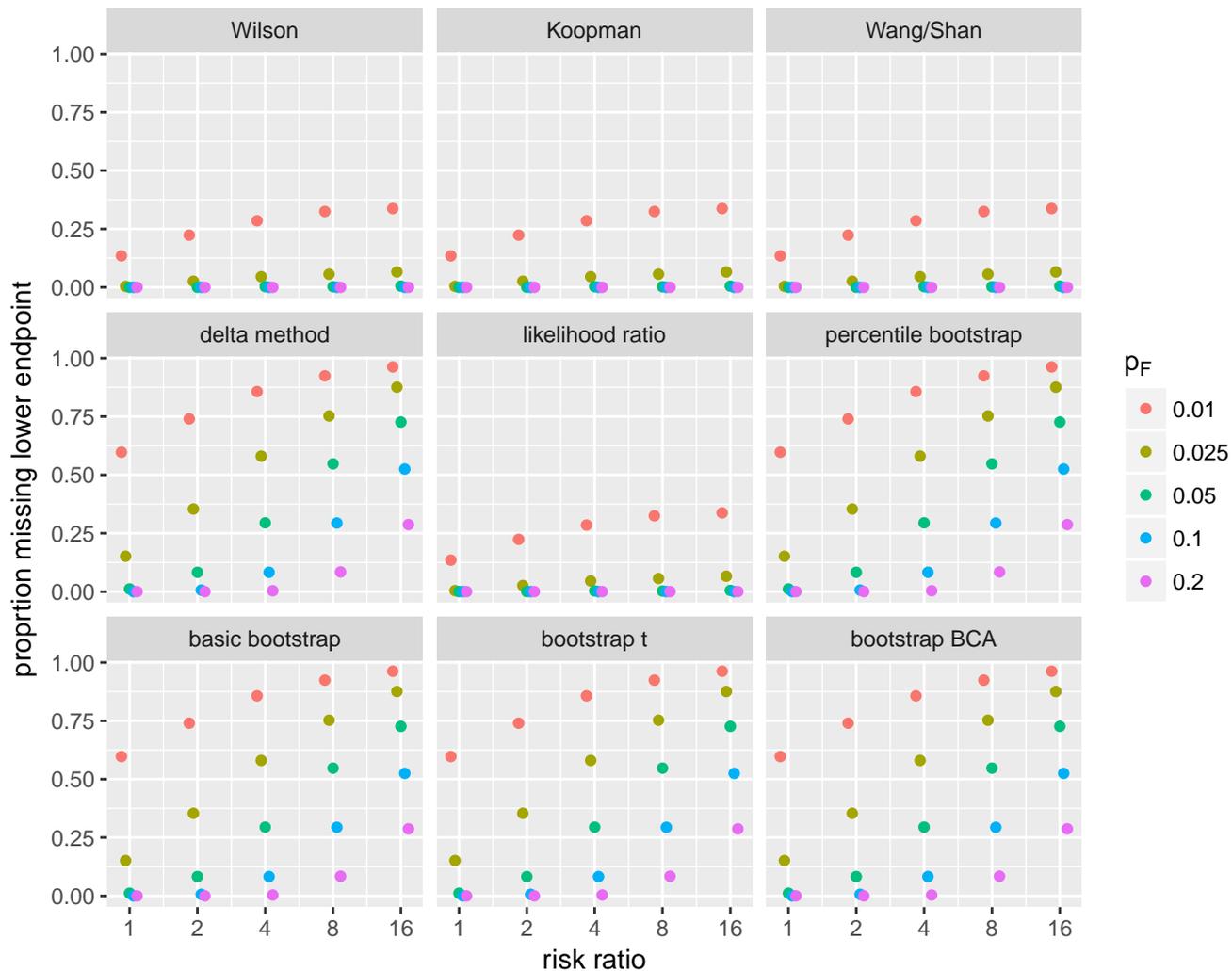}

\caption{Proportion of simulated datasets for which lower confidence bound
could not be calculated for $n=100$. For the Wilson, Koopman, Wang-Shan,
and likelihood ratio methods this occurs only when no events occur
in both scenarios.}
\label{fig:missing_low}
\end{figure}

Finally, Figure \ref{fig:missing_low} shows the proportion of simulated
datasets for which a lower confidence bound could not be calculated,
illustrating how commonly the bootstrap and delta methods fail, particularly
for larger values of RR and lower values of $p_{F}$, which correspond
to scenarios with fewer events.

Figures \ref{fig:coverage_low-n400}-\ref{fig:missing_low-n400} in
the Supplemental Material A show coverage results for $n=400$. In
general these results are similar to those for $n=100$, but for $n=400$
we see that coverage probability is generally closer to 95\%, as we
would expect with increasing sample size. Results for $n\in\{25,50\}$
were produced but are not shown; the results are similar quantitatively
in terms of coverage and qualitatively in terms of the median bound
values, but as expected show smaller lower bounds and higher proportions
of datasets in which the bound(s) could not be calculated. 

In summary, the bootstrap methods perform poorly. The LR method, which
can be used both when analyzing binomial counts and when using EVA,
provides some advantages, in particular its ability to provide intervals
in most situations, including when $\hat{p}_{C}=0$. In contrast,
the bootstrap intervals require $\hat{p}_{C}>0$ and do not provide
meaningful intervals when too many of the bootstrap samples result
in $\hat{p}_{C}^{(i)}=0$. 

However, the LR interval shows undercoverage; in contrast, if one
is using binomial counting, then the Koopman and Wang-Shan methods
are possibilities and avoid the undercoverage of the LR method. However,
the upper endpoint of Koopman shows some undercoverage and both methods
can be too conservative, leading to overly-long intervals. Thus there
is a tradeoff here, and a user may wish to consider how conservative
they wish to be in their analysis. The Wilson method performs similarly
to the Koopman method for the lower endpoint but has extreme undercoverage
for the upper endpoint. The Koopman method is easy to compute, although
the endpoints for Wang-Shan (whose calculation is computationally-intensive)
can be pre-computed and saved as a lookup table.

\section{Case study: Texas drought/heatwave}

\label{sec:example}

\subsection{The event}

Our case study focuses on the heat and dryness over the US state of
Texas in the summer of 2011, the hottest and driest (in terms of precipitation
deficit) on record dating back over a century \citep{Rupp_Mote_2012,Hoer_etal_2013}.
\citet{Rupp_Mote_2012} performed a probabilistic event attribution
analysis, but with counterfactual conditions estimated from previous
years with similar anomalous temperature patterns in the Pacific Ocean
(but cooller ocean temperatures generally). They concluded that anthropogenic
emissions had increased the chance of the anomalously low rainfall
and especially of the anomalous heat over Texas. While \citet{Hoer_etal_2013}
also found evidence for an anthrogenic contribution to the anomalous
heat, they did not see evidence of a contribution to the precipitation
deficit. Revealing possible inconsistencies in methods, however, these
estimates based on climate models are at odds with the lack of an
observed long-term summer warming over Texas \citep{Ston_etal_2013,Hoer_etal_2013}.
In this case study, we follow \citet{Rupp_Mote_2012} in considering
March-August growing season temperature and rainfall averaged over
the state of Texas, though the methods discussed in this work apply
to longer or shorter periods.

\subsection{Data and methods}

We use data from CAM5.1 contributed to the C20C+ Detection and Attribution
Project \citep{Foll_etal_2014}. CAM5.1 simulates the processes of
the atmosphere and land surface, given prescribed radiative and ocean
surface conditions, run at a spatial resolution of approximately $1^{\circ}$
\citep{Neal_Chen_2012}. Simulations have been run under a factual
scenario of observed boundary conditions (e.g., greenhouse gas concentrations,
aerosol burdens, and sea surface temperatures) and under a counterfactual
scenario with greenhouse gas and other radiative boundary conditions
set to year-1855 values and ocean conditions set to the benchmark
estimate of \citet{Ston_Pall_inprep}. This removes a spatio-temporal
pattern of anthropogenic warming but preserves month-to-month and
year-to-year anomalous variability. We use data from 50 simulations
of both scenarios covering a 1961-2010 reference period, 100 simulations
(including the longer 50) covering the 1997-2010 period, and 400 simulations
(including the longer 100) covering 2011-2013 \citep{Ange_etal_2017}.

While multiple observational datasets are available, we use the CRU-TS-3.22
observationally-based data set \citep{Harr_etal_2014}. We work with
anomalies of both observations and model output against the 1961-2010
period, by subtracting the historical mean for temperature and dividing
by it for precipitation. For the model simulations, the factual scenario
reference using the 50 simulations covering the full 1961-2010 period
is used for estimating anomalies for both of the scenarios.

When using EVA we set the threshold, $u$, as the 90th percentile
for temperature of the values being analyzed (both scenarios). For
precipitation we consider the 20th percentile. 

All code and data are available at \\\url{https://bitbucket.org/lbl-cascade/event_attribution_uq_paper}.
We use the climextRemes R package version 0.2 (also available for
Python through Conda) to estimate risk ratios and uncertainty, in some cases based
on binomial counts and in others on EVA.

\subsection{Uncertainty analysis for temperature}

\begin{figure}
\includegraphics{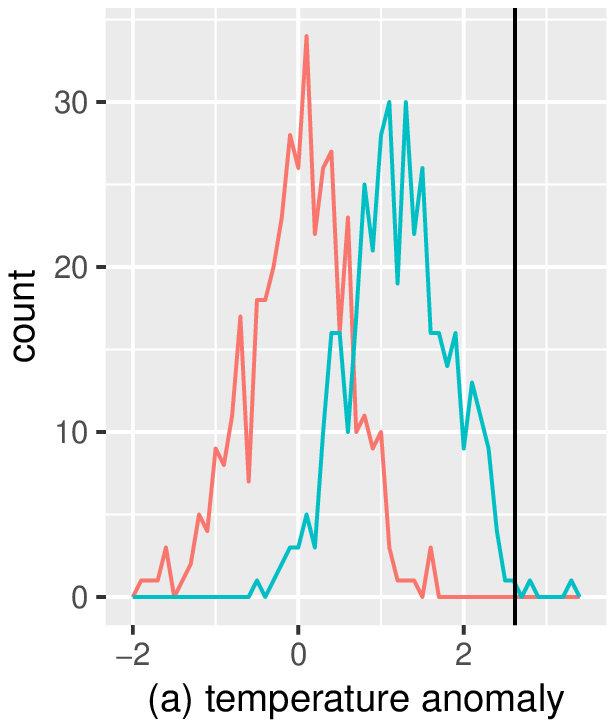}\includegraphics{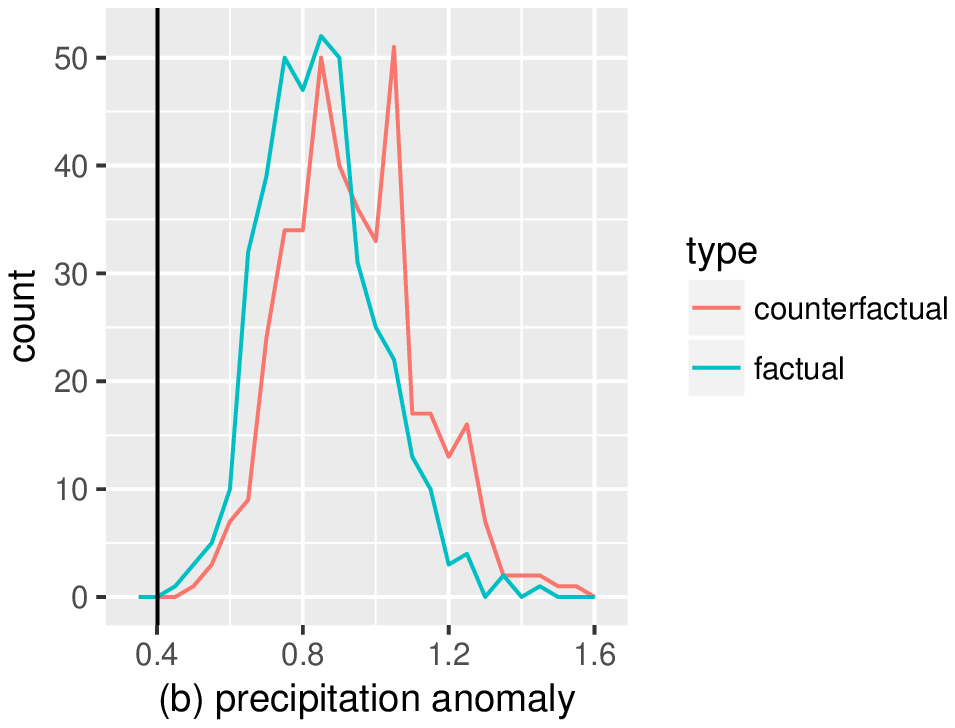}

\caption{Histograms of temperature (degrees Celsius) (a) and precipitation
(b) anomalies for March-August 2011 over Texas from 400-member ensembles,
with actual event indicated by black line. }
\label{fig:hist}
\end{figure}

Fig. \ref{fig:hist}a shows the distribution of temperature anomalies
for 2011 for the two scenarios. 

The actual event of a 2.62$^{\circ}$C anomaly in 2011 is very extreme
relative to both the factual and counterfactual distributions, particularly
so with respect to the counterfactual. For the binomial approach,
the estimated RR is $\infty$. Using the likelihood ratio-based confidence
interval, we have a one-sided 95\% confidence interval (CI) of (1.04,
$\infty$), while using the Koopman method we have (0.74, $\infty$);
both are quite uncertain because of how extreme the event is. Here
EVA provides us with the ability to use the information in the sample
more effectively because the event is so extreme. The RR estimate
is still $\infty$, with $\hat{p}_{A}=0.0067$ and $\hat{p}_{N}=0$,
but the 95\% one-sided CI using the likelihood ratio approach is (12.8,
$\infty$), providing strong evidence for a large RR. 

Next, note that for less extreme event definitions, the event can
be fairly common in the factual scenario and still not observed or
very uncommon in the counterfactual. Given that the event is not extreme
in the factual, EVA is not appropriate for that scenario, and we focus
on the results based on the binomial approach. Table \ref{tab:temp-RR}
shows results for a variety of definitions of the event. Note that
we report a two-sided 90\% CI by considering two one-sided 95\% intervals.

\begin{table}
\caption{RR estimates and 90\% CI for a variety of event definitions based
on binomial count approach. The last three event definitions are based
on the quantiles of the CRU observations.}
\label{tab:temp-RR}

\begin{tabular}{|c|c|c|c|c|}
\hline 
event definition ($\deg\,C$) & Number factual/counterfactual exceedances & $\rrh$ & Koopman CI & LRT CI\tabularnewline
\hline 
\hline 
2.62 (actual event) & 2 / 0 & $\infty$ & (0.74, $\infty$) & (1.04,$\infty$)\tabularnewline
\hline 
2.0 & 43 / 0 & $\infty$ & (16, $\infty$) & (31, $\infty$)\tabularnewline
\hline 
1.5 & 129 / 3 & $43$ & (17, 108) & (19, 133)\tabularnewline
\hline 
1.03 (1 in 20 year) & 245 / 11 & $22$ & (14, 36) & (14, 38)\tabularnewline
\hline 
0.73 (1 in 10 year) & 314 / 40 & $7.9$ & (6.1, 10.1) & (6.2,10.2)\tabularnewline
\hline 
0.43 (1 in 5 year) & 357 / 90 & $4.0$ & (3.4, 4.6) & (3.4,4.7)\tabularnewline
\hline 
\end{tabular}

\end{table}

Note that the factual distribution is shifted so substantially relative
to the counterfactual distribution that the RR estimates are very
large for extreme events and necessarily less so for less extreme
events because $\hat{p}_{F}$ has an upper bound at 1 and as the event
becomes less extreme, $\hat{p}_{C}$ increases and is no longer negligible.

Regardless, the evidence is strong that the probability of an extreme
heatwave is much greater with anthropogenic influence than without,
and for events defined by anomaly values larger than one, we conclude
the risk ratio is at least 10, having accounted for sampling uncertainty.
An important caveat is of course whether the model is able to capture
the climatology of the event of interest, which is outside the scope
of this work.

\subsection{Uncertainty analysis for precipitation}

Fig. \ref{fig:hist}b shows the distribution of precipitation (relative)
anomalies for 2011 for the two scenarios. For the actual event, an
anomaly value of 0.40 (i.e., 40\% as much precipitation as the historical
mean), we have zero exceedances in both scenarios. Furthermore EVA
is of limited use as we get $\hat{p}_{F}=0.000088$ and $\hat{p}_{C}=0$,
with high uncertainty: the likelihood-ratio based lower bound for
RR is less than 0.01. 

\begin{figure}
\includegraphics{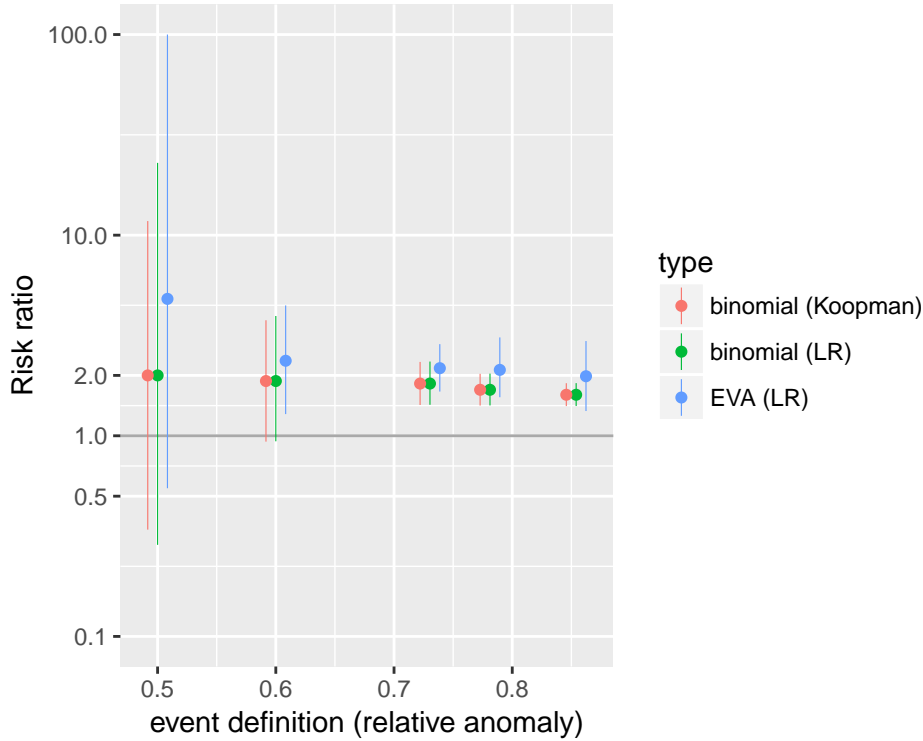}

\caption{Estimated risk ratio and 90\% confidence intervals for both binomial
counts and EVA for various definitions of the event in terms of March-August
precipitation anomaly over Texas (so lower values are more extreme).
The three event definitions to the right are the 1 in 20, 1 in 10
and 1 in 5 year events based on the CRU observations. The upper bound
for EVA for the event of 0.5 is greater than 100 but truncated at
100 for plotting purposes. Note that for less extreme event definitions,
EVA is less appropriate as EVA relies on the cutoff being in the tail
of the distribution.}
\label{fig:precip-estimates}
\end{figure}

Given this, we consider several values for the event definition that
are less extreme than the actual drought. Fig. \ref{fig:precip-estimates}
shows results from both the binomial count approach and EVA. In general
we see evidence for a RR greater than one, with our best estimate
of a RR of about two at a variety of event definitions. However the
uncertainty in the lower bound at more extreme event definitions limits
our ability to make a more robust attribution statement. 

\section{Discussion and recommendations\label{sec:discussion}}

We have presented a statistical framework for estimating the risk
ratio to quantify anthropogenic influence on extreme events, focusing
on proposing, implementing, and evaluating standard frequentist statistical
methods for accounting for sampling uncertainty in event attribution.
Our methodological recommendations are as follows:
\begin{enumerate}
\item When using binomial counting, we recommend either the Koopman or Wang-Shan
methods. If one is interested in the upper endpoint as well as the
lower endpoint for small sample sizes, then Koopman does not preserve
coverage probability, and we suggest Wang-Shan. These methods can
be conservative, so the likelihood ratio approach might also be considered
despite it producing intervals that can be too short.
\item When using EVA, only the normal-theory, likelihood ratio, and bootstrap
methods are feasible. We recommend the likelihood ratio approach to
calculate confidence intervals for the RR. This approach can provide
a CI even when the estimate of the RR is infinity and avoids the problem
that the bootstrap can fail to varying degrees when bootstrapped values
of the RR estimate cannot be calculated or are infinity.
\item With the likelihood ratio approach, confidence intervals can be too
short in a variety of circumstances and should be interpreted cautiously.
\item While appealing in its simplicity, bootstrap methods can perform poorly
for quantifying uncertainty in RR, particularly when either $p_{F}$
or $p_{C}$ are near zero.
\item Bootstrapping provides a methodology to calculate confidence intervals,
not Bayesian probability statements/intervals about RR. Unless an
explicit Bayesian analysis is done, researchers should provide confidence
intervals as their measure of uncertainty and avoid plotting the bootstrap
distribution, as it represents the sampling distribution for $\rrh$
not the distribution of RR.
\item Although we have not investigated its performance in the simulation
study, EVA may provide estimates that use the data more efficiently
than binomial counting in some cases, particularly when sample sizes
are not too small and the event definition is fairly extreme under
both scenarios. In this situation, the small counts of extreme events
lead to high uncertainty for the binomial counting approach, but EVA
can borrow information from data values below the event definition
cutoff. However, with small sample sizes, EVA suffers from having
few observations with which to fit the distribution and often from
difficulties in numerical optimization. 
\item EVA is not appropriate if the event is not extreme in the scenario
being analyzed (e.g., often the case for cold events in a natural
counterfactual scenario). Furthermore simple binomial counts are effective
and straightforward as sample sizes increase provided the event is
not too rare in at least one of the scenarios. EVA and binomial counting
could be used in the same analysis when the event is rare in one scenario
and not the other. However, in this case it is clear that the RR is
far from zero and there would be limited benefit from reducing uncertainty
in the probability for the scenario in which the event is rare, so
analysis using binomial counting for both scenarios should be very
effective.
\end{enumerate}
Bayesian methods can also be useful and can be used to account for
sampling uncertainty. Perhaps most usefully, a Bayesian perspective
is probably the only option for accounting for uncertainty from uncertain
boundary conditions (including uncertain counterfactual conditions
in conditional attribution studies) and model parametric and structural
uncertainty. For example, if one has an ensemble of model simulations
based on varying the parameters of the climate model, this is most
naturally seen as having drawn from a (prior) distribution over model
parameters. This is naturally viewed in a Bayesian context but is
difficult to conceptualize in a frequentist statistics framework as
variation in the data that one might observe under the hypothetical
of repeating an experiment. Similarly, the use of multiple climate
models and the possibility of an ensemble of simulations of possible
aerosol forcing or other forcings could be considered in a Bayesian
framework \citep[e.g.,][]{Smit_etal_2009}. Critically, unlike with
sampling variability, additional simulations do not reduce uncertainty
from these sources or give results that converge in a frequentist
statistical context to the true RR. In fact, from a statistical perspective
one might characterize these uncertainties as bias rather than variance.
Given that conditioning on the ensemble output has limitations in
terms of quantifying these uncertainties, any statistical treatment
of these uncertainties is likely to represent some form of sensitivity
analysis or subjective Bayesian analysis. Consideration of how to
quantify parametric and structural uncertainty is an active area of
climate research generally \citep[e.g.,][]{Knut_etal_2010}.

One common question in event attribution analyses is how to define
the event, which is often based on choosing the cutoff beyond which
an 'extreme event' is considered to have occurred. When the motivation
for a study is damage to society, one may be able to choose a specific
cutoff (or small range of cutoffs) to use. If analysis is motivated
by an event occurring and the cutoff is determined based on observations,
one might consider the definition to be a source of uncertainty. However,
our perspective (originally developed in \citet{Jeon_etal_2016})
is that this is most naturally considered as a sensitivity analysis,
reporting the RR and uncertainty for a variety of cutoff values, as
done in our case study and in \citet{Ange_etal_2017} and \citet{Pall_etal_2017}.
In some cases, such as temperature in our Texas example, while the
results vary somewhat with the event definition, the lower bound of
the confidence interval provides evidence for a robust attribution
statement in light of uncertainty.

\section*{Acknowledgments}

We thank Oliver Angélil for contributing the C20C+ D\&A model simulations
used in the analysis of the Texas event and Weizhen Wang for code
to implement the method of \citet{Wang_Shan_2015}. This research
was supported by the Director, Office of Science, Office of Biological
and Environmental Research of the U.S. Department of Energy under
Contract No. DE-AC02-05CH11231 as part of their Regional \& Global
Climate Modeling (RGCM) Program within the Calibrated And Systematic
Characterization Attribution and Detection of Extremes Scientific
Focus Area (CASCADE SFA).

\bibliographystyle{/accounts/gen/vis/paciorek/latex/RSSstylefile/Chicago}

\bibliography{eventAttrPaper}

\part*{Supplemental Material}

\section*{A Additional simulation study results for larger sample size\label{app:more_sims}}

\begin{figure}
\includegraphics{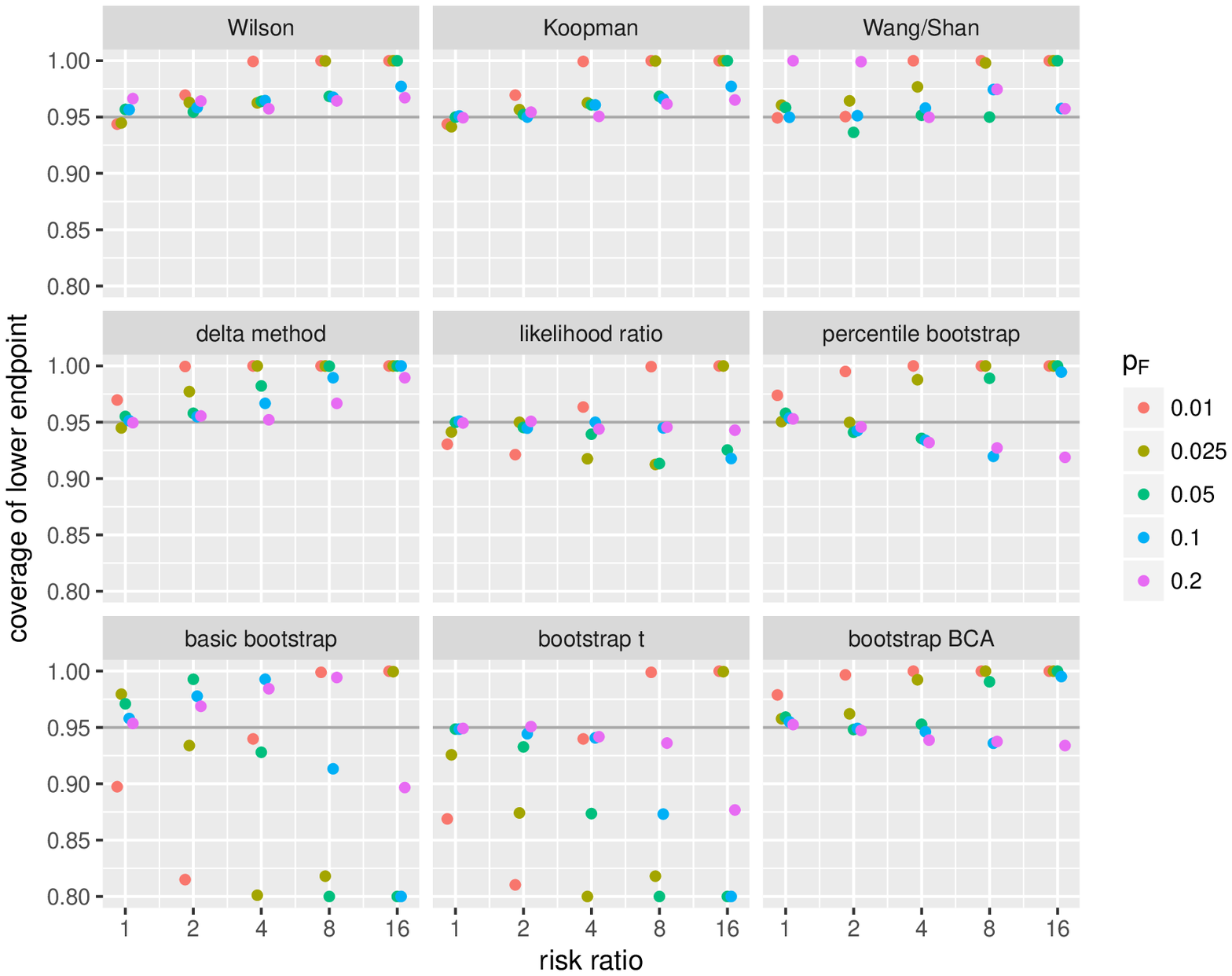}

\caption{Coverage probability of 95\% lower confidence bound for $n=400$ for
various methods and values of RR and $p_{F}$. Values of 0.95 are
optimal, while values less than 0.95 indicate undercoverage and values
greater than 0.95 indicate conservativeness (overcoverage). Values
lower than 0.8 are set to 0.8 for display purposes. Simulations in
which the lower bound could not be computed were excluded.}
\label{fig:coverage_low-n400}
\end{figure}

\begin{figure}
\includegraphics{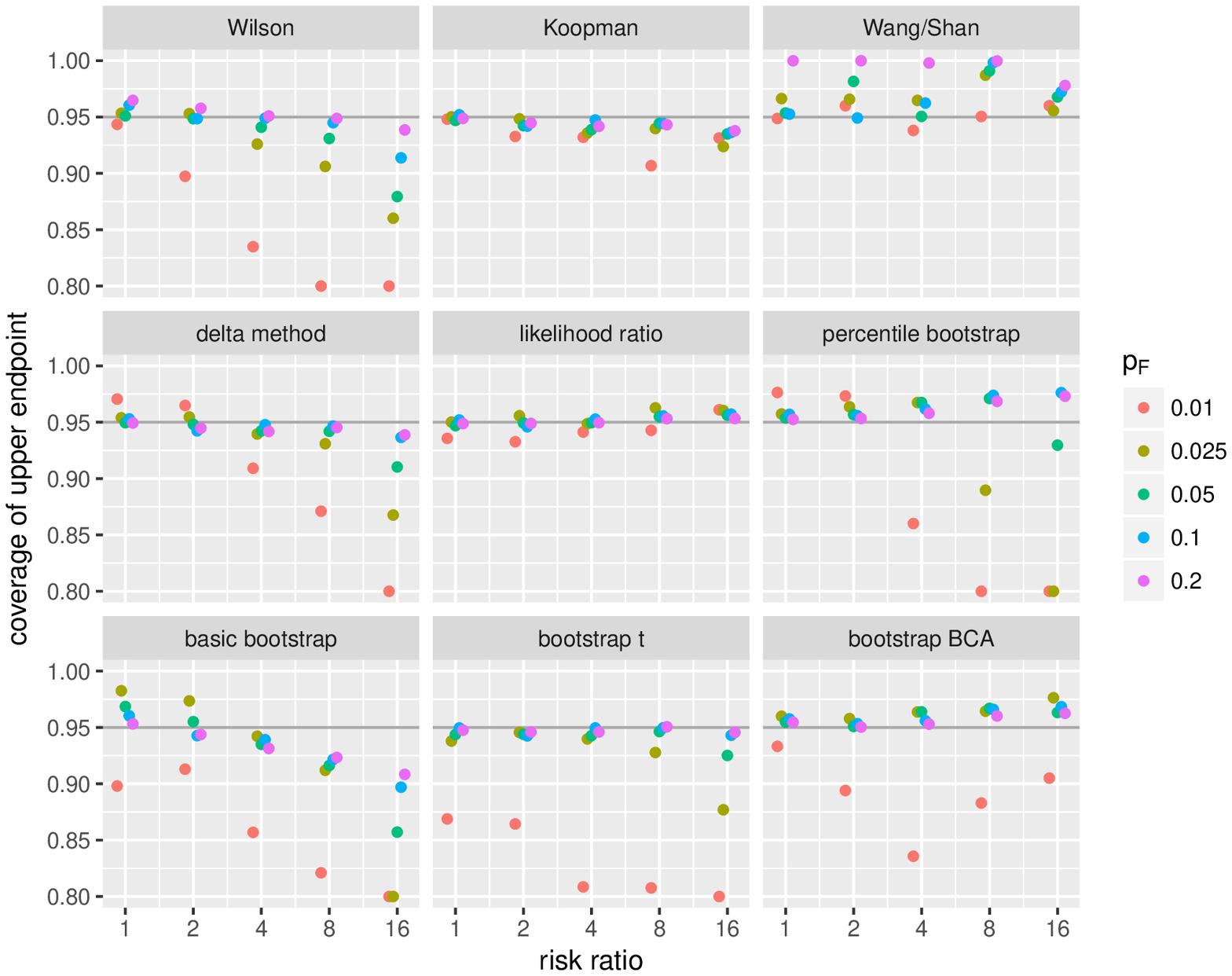}

\caption{Coverage probability of 95\% upper confidence bound for $n=400$ for
various methods and values of RR and $p_{F}$. Values of 0.95 are
optimal, while values less than 0.95 indicate undercoverage and values
greater than 0.95 indicate conservativeness (overcoverage). Values
lower than 0.8 are set to 0.8 for display purposes. Simulations in
which the upper bound could not be computed were excluded.}
\label{fig:coverage_high-n400}
\end{figure}

\begin{figure}
\includegraphics{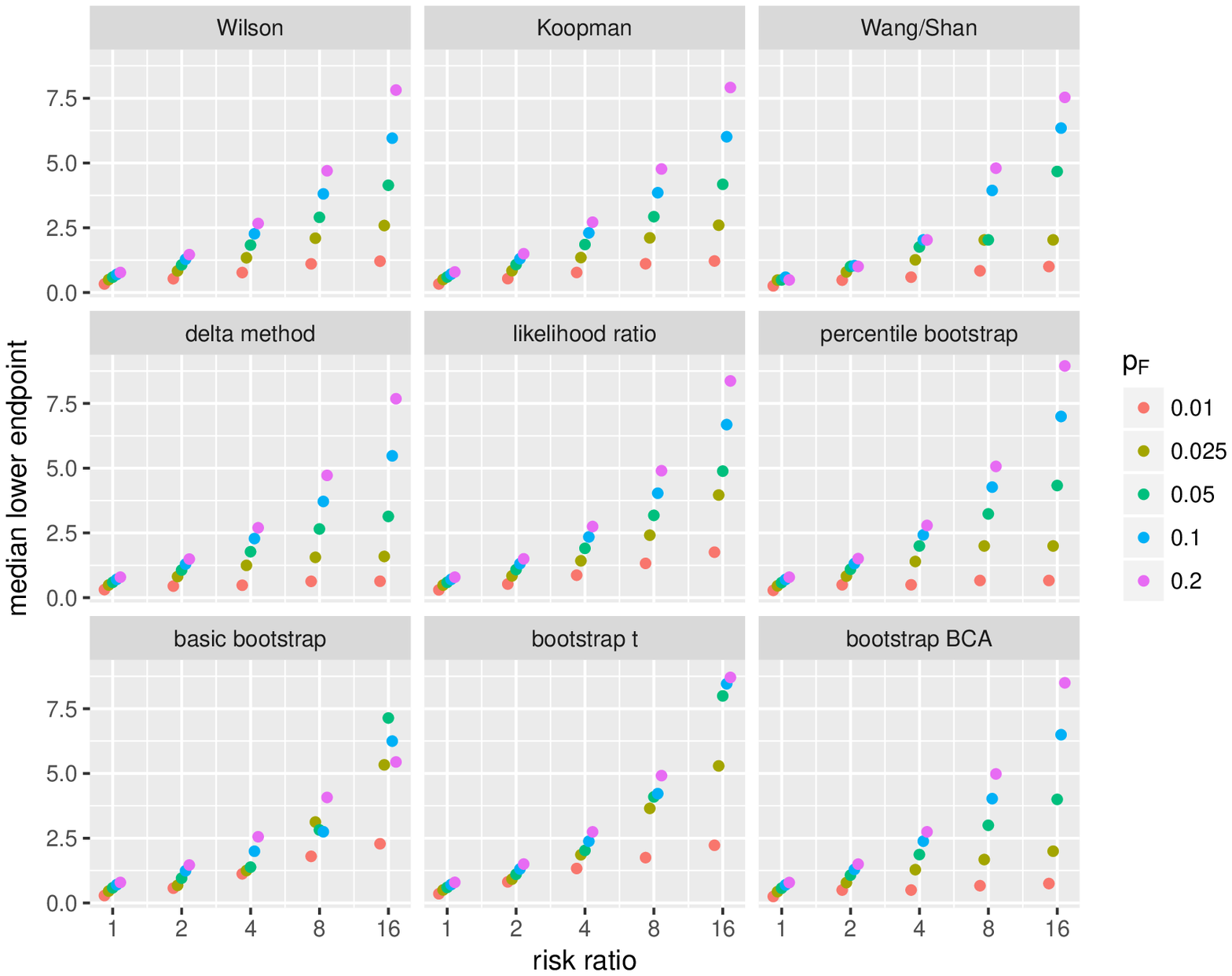}

\caption{Median value of lower confidence bound for $n=400$ for various methods
and values of RR and $p_{F}$. Higher values are better as they correspond
to shorter confidence intervals. Simulations in which the lower bound
could not be computed were excluded.}
\label{fig:value_low-n400}
\end{figure}

\begin{figure}
\includegraphics{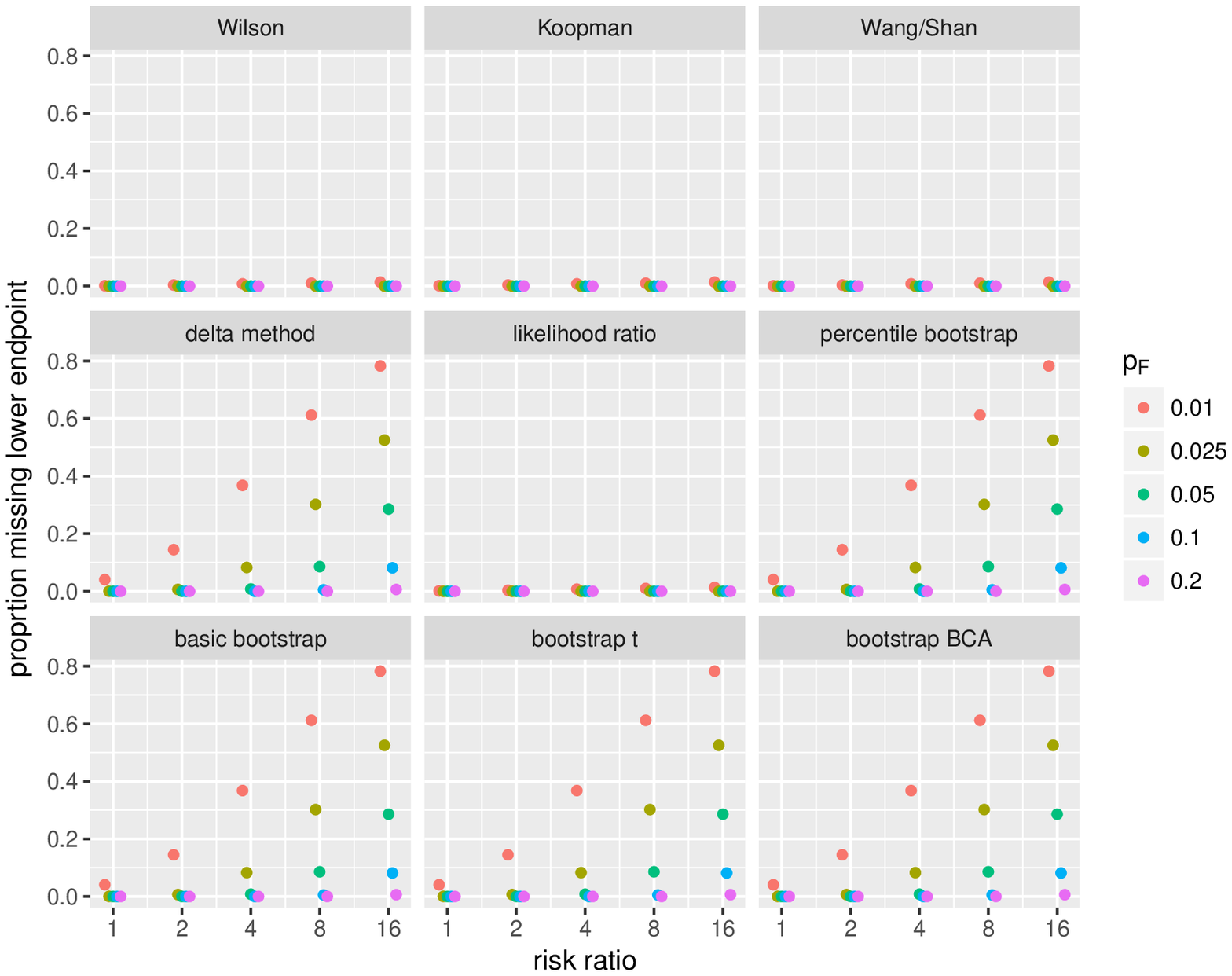}

\caption{Proportion of simulated datasets for which lower confidence bound
could not be calculated for $n=400$. For the Wilson, Koopman, Wang-Shan,
and likelihood ratio methods this occurs only when no events occur
in both scenarios.}
\label{fig:missing_low-n400}
\end{figure}

Figures \ref{fig:coverage_low-n400}-\ref{fig:missing_low-n400} present
simulation results for a larger ($n=400$) sample size, for comparison
with the results in the main body for $n=100$.

\section*{B Accounting for uncertainty in simulations without ocean dynamics\label{app:AGCMs}}

\subsection*{B.1 Framework and additional sources of uncertainty}

In this section we discuss the complications induced when doing event
attribution using atmosphere-land-only model simulations \citep[e.g.,][]{Pall_etal_2011}.

First, we highlight the conditional nature of the event attribution
analyses that use atmosphere-land GCMs and prescribed sea surface
temperatures (SSTs). This approach retains anomalous SST features,
such as the existence of an observed El Niño event, in both the factual
and counterfactual scenarios. Thus it conditions on the SST state
and so provides a probability ratio estimate that is conditional on
the observed SST pattern and the estimated difference in magnitude
of the SSTs between the two scenarios. This means that the framework
ignores the possibility that anthropogenic factors influence the likelihood
of a given SST state. I.e., we are quantifying the ratio of \emph{conditional
}probabilities,
\[
\frac{p_{F}(R>c|S=s_{F})}{p_{C}(R>c|S=s_{C})},
\]
for particular SST states $s_{F}$ and $s_{C}$. 

In this conditional context, there are two additional sources of uncertainty
in event attribution analyses:
\begin{itemize}
\item \textbf{longer-term internal variability of the system}: If we are
sampling from simulations of atmospheric-land GCMs, then variability
in the longer-term state of the earth system at yearly and multi-year
time scales is not included in a simple initial condition ensemble
of simulations in the same way that short-term variability is. This
distinction does not exist if we are sampling from long simulations
of GCMs that also simulate ocean processes.
\item \textbf{ocean and ice counterfactual boundary conditions}: the lack
of ocean and ice dynamics adds an additional aspect of uncertainty
-- the uncertainty of the SST and sea ice boundary conditions under
the counterfactual scenario. \citet{Pall_etal_2011} took one approach
to try to address this, but treatment of this uncertainty shares similarities
to the uncertainties induced by use of a model discussed in Section
\ref{sub:Sources-of-uncertainty}. 
\end{itemize}

\subsection*{B.2 Estimating a partly unconditional risk ratio from conditional
analyses}

\label{par:internal-var-estimator}

If we want an unconditional risk ratio, we seek to calculate: 
\[
\frac{p_{F}(R>c)}{p_{C}(R>c)}=\frac{\int p_{F}(R>c|s)f_{F}(s)ds}{\int p_{C}(R>c|s)f_{C}(s)ds},
\]
which averages conditional probabilities over the marginal distribution
of the state of the system ($S$) (i.e., the SST state), under the
factual, $f_{F}(s)$, and counterfactual, $f_{C}(s)$, scenarios.
To estimate the integrals using Monte Carlo methods we need to be
able to draw from the SST distributions. 

Fortunately, the presence of multiple years in each simulation of
an atmosphere-land GCM provides some ability to account for internal
variability. Note that there remains some conditionality though, in
the sense that the sample of SST states is still defined as identical
for both scenarios. Here we discuss a basic frequentist analysis suitable
for this scenario. An alternative Bayesian statistical strategy is
developed in \citet{Riss_etal_2017}. 

The basic approach is to assume that the SST state is approximately
independent from one year to the next such that we can treat the years
as a simple random sample over the internal variability of the system.
Given an initial condition ensemble, we can consider a probability,
$p_{t}$, for year $t$, that conditions on the fixed SST state:

\[
p_{t}=P(R_{t}>c)=E(I(R_{t}>c))=\int I(R(w)>c)f_{t}(w)dw.
\]
Here $f_{t}(w)$ is the probability density of weather for a given
year (i.e., in a given setting of the internal variability of the
system). We can estimate $p_{t}$ by $\hat{p}_{t}$ nonparametrically
using a standard Monte Carlo estimator applied to the ensemble output
for year $t$ as in (\ref{eq:MC_estimator}) or using techniques discussed
in Section \ref{sec:Estimating-event-probabilities}.

If we are interested in a single probability, $p$, that represents
the probability of the event averaging over internal variability (we
would carry out these calculations separately for the factual and
counterfactual scenarios), then our quantity of interest is

\begin{eqnarray*}
p & = & P(R>c)=E(I(R>c))=\int I(R(w)>c)f(w)dw\\
 & = & \int I(R(w)>c)f(w|s)f(s)dwds
\end{eqnarray*}
where $S$ is a random variable representing the state of the system
and $f(s)$ is the probability density of $S$. Here the distribution
of weather, $W$, can be decomposed into the conditional distribution
of $W$ given the state of the system, $S$, multiplied by the distribution
of the state of the system. The quantity just above can also be expressed
as
\begin{eqnarray*}
p & = & \int\left\{ \int I(R(w)>c)f(w|s)dw\right\} f(s)ds\\
 & = & \int p_{s}f(s)ds.
\end{eqnarray*}
$p_{s}$ is equivalent to $p_{t}$ above when time and the state of
the system are considered to be equivalent, and hence we can think
of $p$ as being the average of $p_{t}$ over the distribution of
the state of the system.

A Monte Carlo estimate of $p$, averaging over internal variability
as represented by the multiple years, $n_{t}$, can be obtained as
follows. 

\begin{eqnarray}
\hat{p} & = & \frac{1}{n_{t}}\sum_{t=1}^{n_{t}}\hat{p}_{t}=\frac{1}{n_{t}}\sum_{t=1}^{n_{t}}\frac{1}{n_{w}}\sum_{i=1}^{n_{w}}I(R(w_{it})>c)\label{eq:phat-internalvar}\\
 & = & \frac{1}{n_{t}n_{w}}\sum_{t=1}^{n_{t}}\sum_{i=1}^{n_{w}}I(R(w_{it})>c)\nonumber 
\end{eqnarray}
which is simply the proportion of exceedances of the cutoff across
the full ensemble by time sample.

Note that this approach assumes that the sample of time reasonably
approximates a simple random sample of the internal variability of
the system. At the least, this requires a sufficient number of years
to capture the different system states, such as ENSO regimes. When
this is not the case, we recommend reporting estimates of $\hat{p}_{t}$
(and the resulting $\rrh_{t}$) for the various years as a sensitivity
analysis rather than reporting a single time-averaged estimate, $\hat{p}$
(and resulting $\rrh$).

\subsection*{B.3 Confidence intervals accounting for internal variability}

\label{sub:internal_var_ci}

When considering methods for finding confidence intervals in this
conditional context of fixed ocean surface conditions, we must account
for the fact that our estimator of $p$ is not based on a simple random
sample of $\{W,V\}$ from the joint distribution. Rather we sample
the internal variability (with year as the proxy) and then sample
$n_{w}$ weather samples, with all $n_{w}$ samples sharing the same
state of the internal variability. This complicates construction of
confidence intervals, and we cannot directly use the methods described
so far except when estimating separate RR values for different time
points. 

One can derive an analogous normal-theory confidence interval to that
discussed in Section \ref{sub:Normal-theory-confidence-interva} by
first calculating the variance of the time-averaged estimator $\hat{p}$
, 
\[
\mbox{Var}(\hat{p})=\frac{1}{n_{t}^{2}}\sum_{t=1}^{n_{t}}\frac{p_{t}(1-p_{t})}{n_{w}},
\]
and then applying the delta method to the log RR calculated based
on the time-averaged estimates, $\hat{p}_{F}$ and $\hat{p}_{C}$. 

The likelihood ratio-based method is not straightforward to apply
when estimating $p$ as an average of time-specific $p_{t}$ values,
because we would need to find the constrained maximum likelihood estimators
under the complicated constraint 
\[
\mbox{RR}_{0}=\frac{\sum_{t=1}^{n_{t}}p_{F,t}}{\sum_{t=1}^{n_{t}}p_{C,t}}.
\]

A bootstrap sample consistent with the actual sampling mechanism is
to resample $n_{t}$ years with replacement and to resample $n_{w}$
ensemble members with replacement. One then uses the same $n_{t}$
resampled years from each of the $n_{w}$ resampled ensemble members.
Note that our resampling procedure will show a lot of variability
in $\rrh$ from one resampled dataset to the next when there is a
lot of variability across years and the number of years simulated
by the model is limited. If we did a bootstrap where we independently
resampled pairs from the ensemble members and years, we would see
less variability than we should as we would not respect the dependence
structure of the original Monte Carlo procedure, namely that all ensemble
members share the same set of years. As a side note, an equivalent
approach would be to resample $n_{w}$ ensemble members and to calculate
$\hat{p}_{t}$ for $t=1,\ldots,n_{t}$ for the $n_{t}$ years of model
simulations and then resample with replacement from $\{\hat{p}_{1},\ldots,\hat{p}_{n_{t}}\}$.
Finally, since the SST conditions under the two scenarios are related,
one should use the same resampling of years under the two scenarios.

Given that we are averaging over multiple years, it is less likely
that the normal-theory and bootstrap intervals cannot be calculated,
but this can still occur.

\subsection*{B.4 Case study}

\begin{figure}
\includegraphics{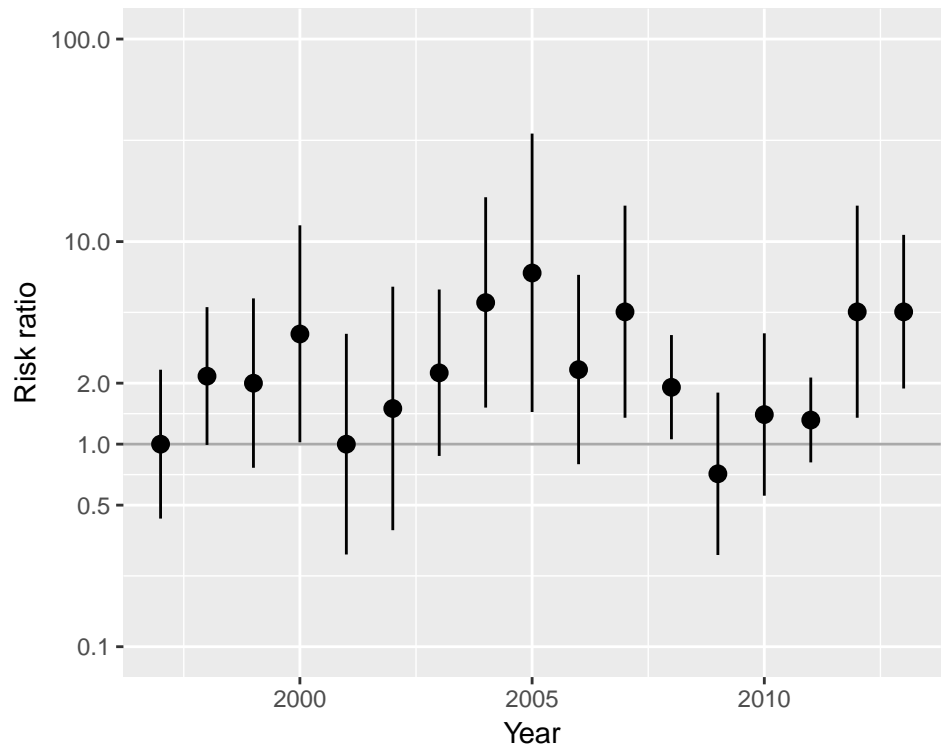}

\caption{Estimated risk ratios and 90\% confidence intervals for March-August
precipitation over Texas by year based on binomial counts with Koopman-based
confidence intervals.}
\label{fig:time-vary}
\end{figure}

Here we extend the precipitation portion of the case study described
in Section \ref{sec:example} to account for internal variability
in conditional analyses. Using the period 1997-2013, for which we
have at least 100 ensemble members available in each year, we consider
accounting for internal variability, with the extreme event defined
as the 1 in 20 year event based on the 1961-2010 CRU data, which is
282 mm. In Fig. \ref{fig:time-vary}, we show the time-varying estimates,
$\rrh_{t}$, with uncertainty based on the binomial count approach
using the Koopman method to estimate confidence intervals. Results
using EVA (not shown) are fairly similar, though the interval endpoints
differ somewhat from the binomial-based intervals. Unfortunately for
two of the years, the EVA likelihood maximization does not converge,
failing to give a result. 

Texas drought tends to be associated with La Ni\~na conditions in
the tropical Pacific, as occurred in 2011 as well as in 1999, 2000,
and 2008; these years do not stand out as unusual for the estimated
RR though. Given the uncertainty, it is difficult to say with certainty
whether the RR varies substantially over time, although the results
do suggest that RR is generally greater than one. \citet{Riss_etal_2017}
tackle this problem in depth, developing a Bayesian methodology and
reporting the importance of internal variability on attribution conclusions
for a variety of event definitions and regions across the globe. 

Finally we estimate the overall RR across time, treating the years
as a sample of the internal variability of the system using the methods
described in the previous section. The estimate of the time-averaged
RR is 2.0 with 95\% confidence intervals of (1.6, 2.6) {[}delta method{]},
(1.4, 3.1) {[}percentile bootstrap{]}, (1.3, 2.8) {[}basic bootstrap{]},
(1.2, 2.8) {[}bootstrap-t{]}, and (1.3, 2.9) {[}BCA bootstrap{]}.
While the intervals vary somewhat, the results indicate some confidence
that after accounting for sampling uncertainty and internal variability,
there is evidence that anthropogenic influence has increased the risk
of drought.

\section*{C Extreme value analysis\label{app:eva}}

\subsection*{C.1 Overview}

We will discuss extremes in the right tail of a distribution, but
all of what follows can also be used for extremes in the left tail
by taking the negative of all the values.

Extreme value analysis is often introduced as a methodology for analyzing
maxima (or minima if considering the left tail) of observations within
blocks \citep[Chapter 3]{Cole_2001}, where a block is often taken
as a year in climate science. A standard climatological extreme value
analysis would analyze the yearly maxima of daily data in an observational
dataset of multiple years. Assuming that the maximum is taken over
many individual observations, the distribution of maxima follows a
three-parameter extreme value distribution called the generalized
extreme value (GEV) distribution, whose cumulative distribution function
(CDF) takes the form, 
\begin{equation}
F(y)=\exp\left\{ -\left[1+\xi\left(\frac{y-\mu}{\sigma}\right)^{-1/\xi}\right]\right\} \label{eq:GEV_CDF}
\end{equation}
 for observation $y$ such that $1+\xi(y-\mu)/\sigma>0$. One can
estimate parameters of the distribution (the location parameter, $\mu$;
the scale parameter, $\sigma;$ and the shape parameter, $\xi$) using
maximum likelihood (or in some cases using a technique called L-moments).
Given parameter estimates, one can estimate return periods and return
values using $F(y)$. For example the probability of an observation
exceeding a value $z$ (the return value) in a given block of observations
(e.g., a year) is $p=1-F(z)$, namely one minus the probability that
the maximum in the block does not exceed $z$. To estimate the probability
of exceeding a specific cutoff $c$ in our earlier notation, we simply
set $z=c$. The return period is $T=1/p$. The return value for a
return period $T$ is found using the inverse CDF, $z=\mu+\sigma\left(\frac{1}{\xi}(-1-\log(1-\frac{1}{T}))\right)^{-\xi}$. 

An alternative approach to EVA that we favor is the peaks-over-threshold
approach based on a point process (PP) model \citep[Chapter 7]{Cole_2001}.
Instead of using block maxima, this approach more effectively uses
the information in the data by making use of all the values exceeding
a high threshold, $u$, often chosen as a high percentile of the empirical
distribution of all the observations, such as the 95th or 99th percentile.
Provided the threshold is sufficiently large, the distribution of
the observations follows a three-parameter extreme value distribution
that is equivalent to the GEV distribution. The likelihood over the
$n_{exc}$ observations that exceed the threshold (out of the total
of $n$ observations taken across all years of data) is
\begin{equation}
\exp\left\{ -n_{t}\left[1+\xi\left(\frac{u-\mu}{\sigma}\right)\right]^{-1/\xi}\right\} \prod_{i=1}^{n_{exc}}\frac{1}{\sigma}\left[1+\xi\left(\frac{y_{i}-\mu}{\sigma}\right)\right]^{-\frac{1}{\xi}-1}.\label{eq:PP_lik}
\end{equation}
When one constructs the likelihood appropriately, as is the case above,
the parameters are equivalent to those of the GEV distribution for
maxima over blocks of size $n/n_{t}$, where $n_{t}$ is the number
of blocks. The number of blocks would generally be the number of years
in climatological observational analysis and the number of total years
across all ensemble members in a model-based analysis. Based on this
equivalence, one can use the GEV CDF (\ref{eq:GEV_CDF}) with the
parameter estimates, $\hat{\theta}=\{\hat{\mu},\hat{\sigma,}\hat{\xi}\}$
obtained from maximizing this likelihood (\ref{eq:PP_lik}) to calculate
return periods and times as discussed in the previous paragraph. In
other words, one uses the point process likelihood to estimate the
parameters and then interprets the parameters as those of the GEV
distribution for maxima. 

All of what we've discussed above can be directly applied to short-term
events such as extreme temperature or precipitation over a day or
limited number of days. However, adjustments need to be made to the
peaks-over-threshold approach to account for daily data not being
independent from day to day, such as the use of declustering techniques
\citep[p. 99]{Cole_2001}.

\subsection*{C.2 Estimating return values with seasonal/yearly data}

It is more complicated when the event of interest occurs over a longer
period such as a season or a year. We are often interested in such
extreme events, e.g., seasonal drought or heat waves. Here we assume
that we have one or a small number of observations in a year, e.g.,
a single summer precipitation value per year. A basic approach using
block maxima would be to group the data into multi-year blocks and
apply GEV analysis to the block maxima. The PP alternative would be
to choose a high threshold and maximize the likelihood. In both cases,
we have much less data than we would with short-term events. Given
this, the greater efficiency of the PP approach compared to the GEV
approach is particularly important. One must be careful in choosing
the threshold, $u$, balancing the desire for more data (leading to
a lower threshold) versus the need for the threshold to be high enough
for the extreme value distribution to be appropriate. One might use
a moderately large percentile of the data, such as the 75th or 80th
percentile. Methods discussed in \citep[Section 4]{Cole_2001} can
be used to guide the choice of threshold but need to recognize that
for seasonal events there are generally few data points available.
In addition, care is needed to find return periods/values that can
be interpreted in terms of probabilities of exceedance in a year in
this context, as follows. 

In the PP likelihood (\ref{eq:PP_lik}), if one takes $n_{t}$ to
be 1 (in essence treating all the years in a single block), then a
probability calculated using the GEV CDF based on $\hat{\theta}$
would be the probability of exceeding a certain value in a period
as long as the entire time period of the observational dataset used.
This is because the equivalent GEV is the distribution of block maxima
for blocks of size $n/n_{t}$. Instead, one can make an adjustment
so that any probability that is calculated is the probability of exceedance
over a year, which can be directly converted to standard return periods.
The adjustment simply sets $n_{t}$ to be the number of years of data
(using the total number of years across all ensemble members in a
model-based analysis). (Note that this has the odd interpretation
for the equivalent GEV model of being the maxima over a single year,
which will be the maximum over a single observation in many cases.)
Alternatively, assuming independence between years, one can convert
between probabilities based on multi-year blocks and single-year blocks
using the relationship $P(m_{b}>z)=1-P(m_{1}<z)^{b}\equiv P(m_{1}>z)=1-P(m_{b}<z)^{1/b}$,
where $m_{b}$ is the maximum over a block of $b$ year(s). 

Finally we discuss how to set the input arguments in various R packages
such that results can be interpreted in terms of probability of exceedance
in a year. In the \emph{climextRemes} package, please see the help
information for the \emph{fit\_pot()} function. If using \emph{fevd()}
in the \emph{extRemes} package, then the \emph{time.units} argument
will be relevant. In particular, if one passes in all of the observations
regardless of whether they exceed the threshold, the implied value
of $n_{t}$ is the number of observations divided by the \emph{npy}
value in the code. When \emph{time.units} is \emph{days} (the default),
$npy=365.25$ and when \emph{time.units} is \emph{years}, $npy=1$.
Similarly with \emph{pp.fit()} in the \emph{ismev} package, the implied
value of $n_{t}$ is the number of observations divided by the argument
\emph{npy}. To accomplish having $n_{t}$ be the number of years or
seasons in the dataset, for \emph{extRemes} this means setting \emph{time.units
}to be \emph{years} (or if one has multiple observations in year,
setting \emph{time.units} to be \emph{x/year} where \emph{x} is the
number of observations in a year). If using the \emph{blocks} argument
for a stationary model in \emph{extRemes}, set \emph{blocks=list(nBlocks
= X)} where \emph{X} is the number of years. And for \emph{ismev}
set \emph{npy} to be 1 (or the number of observations in a year if
greater than one).

\section*{D Statistical details on various confidence interval methods\label{app:stat-details}}

\subsection*{D.1 Normal-theory confidence intervals\label{app:Normal-theory-confidence-interva}}

The Monte Carlo estimator (\ref{eq:MC_estimator}) has a variance,
\begin{equation}
\mbox{Var}(\hat{p})=\frac{\sigma^{2}}{n_{w}}\label{eq:binom-var}
\end{equation}
where $\sigma^{2}=\mbox{Var}(I(R>c))=p(1-p)$ is the standard Bernoulli
variance. Thus the standard error for $\hat{p}$ is $\sqrt{p(1-p)/n_{w}}$
, which can be estimated by 
\[
\widehat{\mbox{se}(\hat{p})}=\sqrt{\frac{\hat{p}(1-\hat{p})}{n_{w}}}.
\]

When $n_{w}$ is sufficiently large, then by the Central Limit Theorem,
$\hat{p}$ is approximately normally distributed and we can calculate
a 90\% confidence interval for $p$ in the usual fashion (plus/minus
1.64 times the standard error). However, particularly when $p$ is
close to zero, we are unlikely to have $n_{w}$ large enough for this
approach to perform well.

\paragraph{Confidence intervals for the RR}

To derive a confidence interval for the log of the risk ratio, which
is a nonlinear function of $p_{F}$ and $p_{C}$, we make use of a
standard statistical method call the delta method (also known as the
method of propagation of errors), which provides valid standard errors
in the limit as the sample size gets large, relying on the central
limit theorem to support the assumption of normality of the estimator.
Consider the log RR expressed as a function of parameters, $\log\mbox{RR}=f(\theta)$
(where $\theta=\{p_{F},p_{C}\}$ in the simple nonparametric binomial
count estimation context). The delta method says that $\mbox{Var}(\log\rrh)=\mbox{Var}(f(\hat{\theta}))\approx\nabla f(\theta)^{\top}\mbox{Cov}(\hat{\theta})\nabla f(\theta)$,
where $\nabla f(\theta)$ is the gradient vector of $f(\cdot)$ with
respect to the elements of $\theta$ and $\mbox{Cov}(\hat{\theta})$
is the covariance matrix of $\hat{\theta}$. Note that if $\hat{\theta}$
is the maximum likelihood estimator, the standard asymptotic estimate
of $\mbox{Cov}(\hat{\theta})$ is based on the inverse of the second
derivative matrix (the Hessian) of the log-likelihood, taken with
respect to the elements of $\theta$, and evaluated at $\hat{\theta}$. 

This can be most simply applied when using the nonparametric binomial
count approach to estimating $p$. In this case, since $\log\mbox{RR}=\log p_{F}-\log p_{C}$
we have $\nabla f(\theta)=\{\frac{1}{p_{F}},-\frac{1}{p_{C}}\}$ ,
which can be estimated by plugging in $\hat{p}_{F}$ and $\hat{p}_{C}$.
$\mbox{Cov}(\hat{\theta})$ is 
\[
\left(\begin{array}{cc}
\mbox{Var}(\hat{p}_{F}) & 0\\
0 & \mbox{Var}(\hat{p}_{C})
\end{array}\right),
\]
where the diagonal elements can be calculated as (\ref{eq:binom-var})
and the off-diagonal elements are 0 because the simulations for the
two scenarios are unrelated. Hence we have 
\[
\widehat{\mbox{se}}(\log\rrh)=\sqrt{\frac{1}{n_{w}}\left(\frac{1-\hat{p}_{F}}{\hat{p}_{F}}+\frac{1-\hat{p}_{C}}{\hat{p}_{C}}\right)},
\]
from which we can derive a confidence interval for $RR$ in similar
fashion to that for $p$ described above.

For situations in which either of the $\hat{p}$ estimates is itself
a nonlinear functional of other estimates, as in the parametric and
EVA approaches, one would apply the delta method directly to log RR
expressed as a function of the original parameters. For example when
using EVA, one has log RR expressed as a function of $\theta=\{\mu_{F},\sigma_{F},\xi_{F},\mu_{C},\sigma_{C},\xi_{C}\}$
where we have estimated GEV parameters for both factual and counterfactual
scenarios that determine $\hat{p}_{F}$ and $\hat{p}_{C}$.

\subsection*{D.2 Inverting a likelihood ratio test\label{app:Inverting-a-likelihood}}

A standard approach to finding a confidence interval is to invert
a test statistic \citep{Case_Berg_2002}. The basic intuition is that
for a hypothesized parameter value, $\theta_{0}$, if we cannot reject
the null hypothesis that $\theta=\theta_{0}$ based on the data, then
that $\theta_{0}$ is a plausible value for the true $\theta$ and
should be included in a confidence interval for $\theta$. A confidence
interval is then constructed by taking all values of $\theta_{0}$
such that a null hypothesis test of $\theta=\theta_{0}$ is not rejected.
For a 90\% interval, the hypothesis test is done at the $\alpha=0.10$
(10\%) significance level.

A standard statistical test that is commonly applicable is a likelihood
ratio test \citep{Case_Berg_2002}, which compares the likelihood
of the data based on the MLE (i.e., the maximized likelihood) to the
likelihood of the data when constraining the parameters to represent
a simpler setting in which the null hypothesis is assumed true (which
in the notation above would be expressed as setting $\theta=\theta_{0}$).
If the null hypothesis were actually true then as the sample size
goes to infinity, twice the log of the ratio of these two likelihoods
has a chi-square distribution with $\nu$ degrees of freedom. $\nu$
is equal to the difference in the number of parameters when comparing
the original parameter space to the restricted space. The hypothesis
that $\theta=\theta_{0}$ is thus rejected when twice the log of the
likelihood ratio exceeds the $1-\alpha$ quantile of the chi-square
distribution. 

Specifically, we are often interested in the plausibility of $p_{F}=p_{C}$
versus the alternative that $p_{F}>p_{C}$, so it would be natural
to derive a one-sided confidence interval, $\mbox{RR}\in(\mbox{RR}_{L},\infty)$,
that gives a lower bound, $\mbox{RR}_{L}$, on the risk ratio. We
illustrate the ideas in the nonparametric binomial count context.
The likelihood ratio test we use here is one where the restricted
parameter space sets $\mbox{RR}=\mbox{RR}_{0}$, so there is one free
parameter, $p_{F}$, with $p_{C}\equiv p_{F}/\mbox{RR}_{0}$. The
unrestricted parameter space has two free parameters ($p_{F}$ and
$p_{C}$) so $\nu=1$ in this setting. For any value $\mbox{RR}_{0}$
we fail to reject the null hypothesis at level $\alpha$ when twice
the log of the likelihood ratio is less than the $1-\alpha$ percentile
of a chi-square distribution with one degree of freedom. The minimum
over the values of $\mbox{RR}_{0}$ that are not rejected gives us
$\mbox{RR}_{L}$. Mathematically, this can be framed as setting $\mbox{RR}_{L}$
as the value of $RR_{0}$ such that 
\begin{equation}
\lambda(RR_{0})=2(\log L(\hat{p}_{F},\hat{p}_{C};y)-\log L(\hat{p}_{F,0},\hat{p}_{C,0};y,\mbox{RR}=\mbox{RR}_{0}))=q_{1-\alpha},\label{eq:LRT-1}
\end{equation}
which can be found via simple one-dimensional optimization or root-finding.
Here $\hat{p}_{F,0}$ and $\hat{p}_{C,0}$ in the second log likelihood
are the maximum likelihood estimates under the constraint that $\mbox{RR}=\mbox{RR}_{0}$,
which can be found in closed form as the solution to a quadratic equation
\citep{Farr_Mann_1990}. $q_{1-\alpha}$ is the $1-\alpha$ quantile
of the chi-square distribution with one degree of freedom. 

While we have derived a lower bound using a one-sided confidence interval,
we could derive an analogous upper bound using a one-sided interval
from the other direction, $\mbox{RR}\in\{0,\mbox{RR}_{U}\}$. Combining
two such 95\% one-sided intervals (i.e., with $\alpha=0.05$, so $q_{1-\alpha}=3.84$)
gives a two-sided 90\% interval.

\paragraph*{Extreme value analysis setting}

The approach above can be generalized to the case where one is estimating
$p_{F}$ and $p_{C}$ by EVA, as proposed in \citet{Jeon_etal_2016}.
In this setting, we have parameters of an extreme value model under
the factual scenario (denoted $\theta_{F}$) and parameters under
the counterfactual ($\theta_{C}$). Each model will generally be a
stationary three-parameter extreme value distribution with $\theta=\{\mu,\sigma,\xi\}$.
We can find the unconstrained likelihood by maximizing with respect
to both $\theta_{C}$ and $\theta_{F}$. For the constrained likelihood,
it will often be most simple to express the constraint, $\mbox{RR}=\mbox{RR}_{0}$
as $p_{C}=p_{F}/\mbox{RR}_{0}$. Under the PP approach, we can constrain
$p_{C}$ by letting $\sigma_{C}$ and $\xi_{C}$ be free parameters
and setting $\mu_{C}\equiv c+\frac{\sigma_{C}}{\xi_{C}}(1-(-\log(1-\frac{p_{F}}{\mbox{RR}_{0}}))^{-\xi_{C}})$
where $c$ is the value of the event for which we are finding the
probability of exceedance. This constrained maximization can be done
in a manner similar to maximizing the PP likelihood, but one needs
to sum the log likelihood terms for both of the scenarios. The numerical
optimization to find $\mbox{RR}_{L}$ can be done as described in
the previous section.

\subsection*{D.3 Inverting hypothesis tests for the ratio of binomial proportions\label{app:Inverting-hypothesis-tests}}

Here we provide an overview of methods from the extensive statistical
literature on confidence intervals for the ratio of two binomial proportions. 

In general for a confidence interval based on inverting a test, one
can frame the inversion of the test based on the p-value associated
with the test. In order to determine the endpoint of a one-sided $1-\alpha$
level interval, one seeks to find the value of the risk ratio for
which the probability of data as or more extreme than observed (i.e.,
the p-value) is equal to $\alpha$. (Two-sided intervals can be found
by constructing two one-sided intervals.) The challenges are (1) determining
the data values that are to be considered 'more extreme', (2) computing
the probability of the data in light of the presence of a second,
nuisance, parameter (namely that there is still a free proportion
after fixing the risk ratio), and (3) carrying out the optimization
necessary to find the value of the risk ratio such that the probability
is $\alpha$. The goal is to find a method for constructing an interval
that keeps the coverage probability greater than or equal to $1-\alpha$,
while giving as short an interval (e.g., as large a lower bound) as
possible. 

\citet{Fage_etal_2015} assess the various methods and find that the
approximate method of \citet{Koop_1984} performs reasonably well
in simulations. They also assess various so-called 'exact' methods
that seek to exactly calculate the p-values used in inverting the
test statistic to derive the interval. Because of the nuisance parameter
($p_{F}$) in addition to the parameter of interest (RR), the exact
methods maximize the p-value over possible values of $p_{F}$. Unfortunately
this maximization can involve maximization of a non-monotonic function
for some data values, which makes the optimization difficult to implement.
Furthermore, by ensuring that coverage is preserved for all values
of the nuisance parameter (even values that are unlikely given the
observed data), intervals based on exact methods can be overly conservative. 

\citet{Wang_Shan_2015} present a new exact approach that hopes to
improve upon existing approaches discussed in \citet{Fage_etal_2015}.
They determine what data values are more extreme than observed using
an inductive process. First they specify a simple, well-defined partial
ordering of possible data values in terms of extremeness (e.g., 5
of 10 exceedances in F with 0 of 10 exceedances in CF is more extreme
than 5 of 10 exceedances in F with 1 of 10 exceedances in CF). However
the logic behind the partial ordering does not describe a full ordering
-- e.g., how does 5/10 vs 0/10 compare to 10/10 vs 1/10? Thus to determine
the full ordering, they sequentially consider all possible orderings
that satisfy the partial ordering and determine the more extreme data
values as those that produce a larger lower bound. For a given ordering,
one can compute the p-value by optimization with respect to the nuisance
parameter, finding the worst case (corresponding to the most conservative
confidence interval) probability over possible values of the nuisance
parameter. The computation then does an optimization to find the smallest
possible value of the risk ratio such that that worst case probability
is equal to $\alpha$. One difficulty is the afore-mentioned non-monotonicity,
which they address using a grid search. The other is that the computations
are time-consuming when the binomial sample sizes are large. 

\end{document}